\documentclass[reprint,twocolumn,secnumarabic,amssymb,aps,pra]{revtex4-2}

\usepackage{titlesec}
\titleformat*{\section}{\large\bfseries}
\titleformat*{\subsection}{\mdseries\bfseries}
\usepackage{graphicx}
\usepackage{float}
\usepackage{wrapfig}
\usepackage{silence}
\usepackage{multirow}
\usepackage{amsmath}
\usepackage{enumitem}
\usepackage{braket}
\usepackage{ulem}
\usepackage{xcolor,colortbl}
\usepackage{makecell}
\usepackage{multirow}
\usepackage[colorlinks,linkcolor=black,citecolor=black,urlcolor=black,filecolor=black]{hyperref}
\makeatletter
\renewcommand*{\fnum@figure}{{\normalfont\bfseries Figure~\thefigure}}
\renewcommand*{\@caption@fignum@sep}{\textbf{ $|$ }}
\makeatother

\newcommand{\GA}{Sample 1}
\newcommand{\CT}{Sample 2}
\newcommand{\NF}{Sample 3}
\newcommand{\PA}{Sample 4}
\newcommand{\MB}{Sample 5}
\newcommand{\NJ}{Sample 6}
\newcommand{\Toluca}{Sample 7}
\newcommand{\bosample}{Sample 8}
\newcommand{\elanasample}{Sample 9}
\newcommand{\papertitle}{Probing spin dynamics on diamond surfaces using a single quantum sensor}

\definecolor{GA_0}{rgb}{0.0, 0.7490196078431373, 1.0}
\definecolor{GA_1_O2}{rgb}{0.6823529411764706,0.7803921568627451,0.9098039215686274}
\definecolor{GA_1_Reset}{rgb}{1.0,0.7333333333333333,0.47058823529411764}
\definecolor{GA_2_O2}{rgb}{0.17254901960784313,0.6274509803921569,0.17254901960784313}
\definecolor{CT_1200}{rgb}{0.8392156862745098,0.15294117647058825,0.1568627450980392}
\definecolor{NF_0}{rgb}{1.0,0.596078431372549,0.5882352941176471}
\definecolor{PA_3_Reset}{rgb}{0.7725490196078432,0.6901960784313725,0.8352941176470589}
\definecolor{MB_1_O2}{rgb}{0.5490196078431373,0.33725490196078434,0.29411764705882354}
\definecolor{MB_2_Reset}{rgb}{0.8901960784313725,0.4666666666666667,0.7607843137254902}
\definecolor{NJ_2_O2}{rgb}{0.9686274509803922,0.7137254901960784,0.8235294117647058}
\definecolor{NJ_2_Reset}{rgb}{0.7803921568627451,0.7803921568627451,0.7803921568627451}
\definecolor{NJ_3_500C}{rgb}{0.7372549019607844,0.7411764705882353,0.13333333333333333}
\definecolor{Toluca_0}{rgb}{0.09019607843137255,0.7450980392156863,0.8117647058823529}
\definecolor{bosample}{rgb}{.122, .467, .706}
\definecolor{elanasample}{rgb}{1, .647, 0}

\begin{document}

\title{\papertitle}

\author{
\normalsize
Bo L. Dwyer$^{1,\ast}$,
Lila V. H. Rodgers$^{2,\ast}$,
Elana K. Urbach$^{1, \ast}$,\\ \normalsize
Dolev Bluvstein$^1$,
Sorawis Sangtawesin$^3$,
Hengyun Zhou$^1$,
Yahia Nassab$^{2\dagger}$,\\ \normalsize
Mattias Fitzpatrick$^{2}$,
Zhiyang Yuan$^{2}$,
Kristiaan De Greve$^{1\ddagger}$,
Eric L. Peterson$^{1}$,\\ \normalsize
Jyh-Pin Chou$^{4}$,
Adam Gali$^{5,6}$,
V. V. Dobrovitski$^7$,
Mikhail D. Lukin$^{1}$,
Nathalie P. de Leon$^{2}$
\\
\it{\normalsize{$^{1}$Department of Physics, Harvard University, Cambridge, MA 02138, USA}}\\
\it{\normalsize{$^{2}$Department of Electrical Engineering, Princeton University, Princeton, NJ, 08544, USA}}\\
\it{\normalsize{$^{3}$School of Physics and Center of Excellence in Advanced Functional Materials, Suranaree University of Technology, Nakhon Ratchasima 30000, Thailand}}\\
\it{\normalsize{$^{4}$Department of Physics, National Changhua University of Education,
Changhua 50007, Taiwan}}\\
\it{\normalsize{$^{5}$Institute for Solid State Physics and Optics, Wigner Research Centre for
Physics, POB 49, H-1525 Budapest, Hungary}}\\
\it{\normalsize{$^{6}$Department of Atomic Physics, Budapest University of Technology,Budafoki út 8., H-1111 Budapest, Hungary}}\\
\it{\normalsize{$^7$QuTech and Kavli Institute of Nanoscience, Delft University of Technology, 2628 CD Delft, The Netherlands}}\\
\normalsize{$^\dagger$ Present address Department of Chemical Engineering, University of Waterloo, Waterloo, ON N2L 3G1, Canada}\\
\normalsize{$^\ddagger$ Present address Inter-university Microelectronics Centre, Imec, Kapeldreef 75, Leuven, Belgium}\\
\normalsize{$^{*}$ These authors contributed equally to this work}
}

\onecolumngrid

\begin{abstract}
    \textbf{Understanding the dynamics of a quantum bit's environment is essential for the realization of practical systems for quantum information processing and metrology. We use single nitrogen-vacancy (NV) centers in diamond to study the dynamics of a disordered spin ensemble at the diamond surface. Specifically, we tune the density of ``dark'' surface spins to interrogate their contribution to the decoherence of shallow NV center spin qubits. When the average surface spin spacing exceeds the NV center depth, we find that the surface spin contribution to the NV center free induction decay can be described by a stretched exponential with variable power $n$. We show that these observations are consistent with a model in which the spatial positions of the surface spins are fixed for each measurement, but some of them reconfigure between measurements. In particular, we observe a depth-dependent critical time associated with a dynamical transition from Gaussian ($n=2$) decay to $n=2/3$, and show that this transition arises from the competition between the small decay contributions of many distant spins and strong coupling to a few proximal spins at the surface. These observations demonstrate the potential of a local sensor for understanding complex systems and elucidate pathways for improving and controlling spin qubits at the surface.}

\end{abstract}

\maketitle
\twocolumngrid
Understanding and controlling the environment of quantum bits is a central challenge in solid-state quantum science and  engineering. Nitrogen-vacancy (NV) color centers in diamond have emerged as a promising platform for numerous applications in quantum sensing and information processing \cite{maze2008nanoscale, cai2013large, schlipf2017molecular}. While bulk NV centers constitute spin qubits with exceptional coherence properties, even under ambient conditions, several applications require shallow NV centers to be placed within nanometers of the diamond surface. In particular, these shallow NV centers can be used for nanoscale sensing applications, enabling detection of individual electron spins \cite{Grotz2011, Grinolds2014}, molecules \cite{Lovchinsky2016a}, or nuclei \cite{Sushkov2014}. However, shallow NV centers exhibit increased decoherence due to surface defects \cite{Myers2014, Romach2015, sangtawesin2019origins}, including ubiquitous ``dark'' unpaired electron spins on the diamond surface \cite{Grotz2011, Grinolds2014}. Prior studies have demonstrated that these dark spins themselves can have long spin relaxation and coherence times, making them potentially useful as ``quantum reporters'' in sensing applications \cite{Sushkov2014, Grinolds2014}. While these surface spins have been observed and studied for several years \cite{Grotz2011, Mamin2012, Grinolds2014, Sushkov2014, sangtawesin2019origins, bluvstein2019extending}, their origin and properties are not well-understood, limiting the potential applications of shallow NV centers to harness them as a resource for sensing and simulation.

Motivated by these considerations, in this work, we systematically investigate the quantum dynamics of over 100 individual shallow NV centers proximal to the diamond surface, using surface treatments to adjust the density of surface spins. We find that the surface spin contribution to the coherence decay, measured across many NV centers in numerous samples, provides insight into the nature of these defects. Moreover, this large dataset permits us to observe a dynamical crossover from Gaussian ($n = 2$) decay to stretched exponential decay ($n < 1$) that is indicative of a dilute two-dimensional spin bath, averaged over many spatial configurations. We show that the crossover time is a generic feature of disordered ensembles, which has broad relevance for other solid-state qubit platforms.

\begin{figure}[ht!]

\includegraphics[width=3.37in]{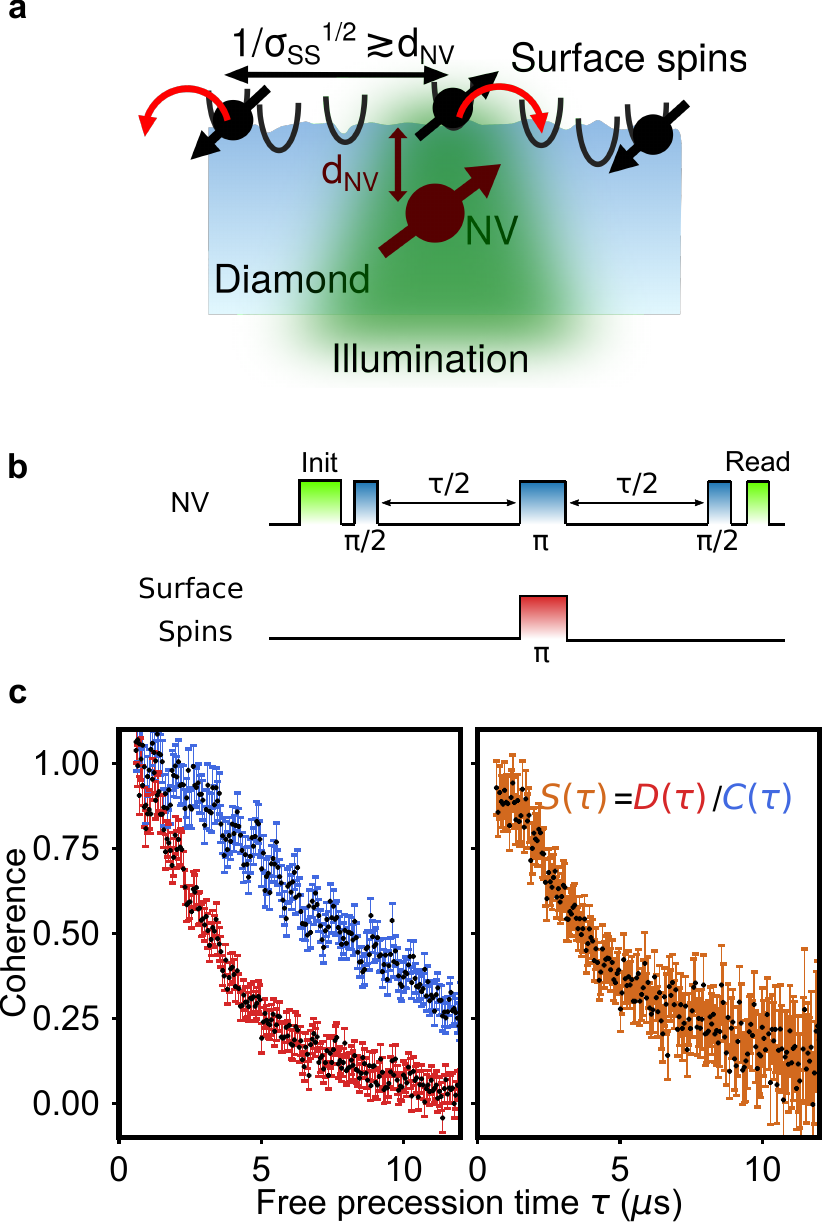}
\caption{\textbf{Double electron-electron resonance of diamond surface spins.} \textbf{a}, Cartoon of experiment, depicting a near surface NV center (dark red arrow) with nearby traps (black wells) that are transiently occupied by spin 1/2 defects (black arrows). A 532 nm laser is used to polarize and readout the NV center. \textbf{b}, Pulse sequence for DEER, where simultaneous microwave $\pi$ pulses are applied to the NV center (blue block) and surface spins (red block) midway through the free precession interval $\tau$. Green blocks indicate NV center spin initialization and readout. \textbf{c}, Spin echo, $C(\tau)$ (blue) and DEER, $D(\tau)$ (red) curves for a representative NV center with coherence oscillations caused by slight magnetic field misalignment (left panel). In order to investigate the DEER signal, $S(\tau)$ (orange) without the influence of decoherence from other sources, $D(\tau)$ is divided by the coherence curve $C(\tau)$ (right panel). Error bars in $\textbf{c}$ are the standard error of the mean due to photon shot noise, propagated point-by-point for the right panel.
}
\label{fig:Fig1}
\end{figure}

Our experiments use NV centers produced by ion implantation \cite{Lovchinsky2016a, sangtawesin2019origins} located at depths $d_\mathrm{NV}$ = 5-20 nm from the surface, under a number of different surface conditions (\hyperref[sec:sampleprocessing]{Supplementary \ref{sec:sampleprocessing}}). We use proton NMR measurements to accurately determine the implanted NV depth \cite{Pham2016}. The surface-spin-NV interaction is studied using double electron-electron resonance (DEER) on a large, random selection of individual NV centers in each sample  \cite{milov1981application}. The DEER pulse sequence, illustrated in Fig.~\ref{fig:Fig1}b, is similar to that of a spin echo, except that microwave $\pi$ pulses are applied to both the NV center spin and the surface spins midway through the free evolution time \cite{Grinolds2014}. This allows for selective interrogation of the flipped spins, while the signal due to non-flipped spins is cancelled to first order by virtue of the echo.

Fig.~\ref{fig:Fig1}c shows an example DEER dataset. In order to remove effects that limit NV center spin coherence in the absence of surface spins, we normalize the DEER decay signal by the coherence signal obtained through a spin echo with no $\pi$ pulse applied to the surface spins \cite{sangtawesin2019origins}. Since the surface spin contribution to decoherence is not canceled by the echo, the resulting decay is equivalent to the free induction decay (FID) caused by the surface spins alone. This normalization technique has the additional benefit of removing oscillations caused by the periodic entanglement with the $^{13}$C nuclear spin bath \cite{childress2006coherent} or slight misalignment between the NV center quantization axis and the externally applied magnetic field \cite{rowan1965electron}. The full data processing and fitting procedures are described in the supplementary materials (\hyperref[sec:fittingprocedures]{Supplementary \ref{sec:fittingprocedures}}).

\begin{figure*}[ht!]
 \includegraphics[width=6.75in]{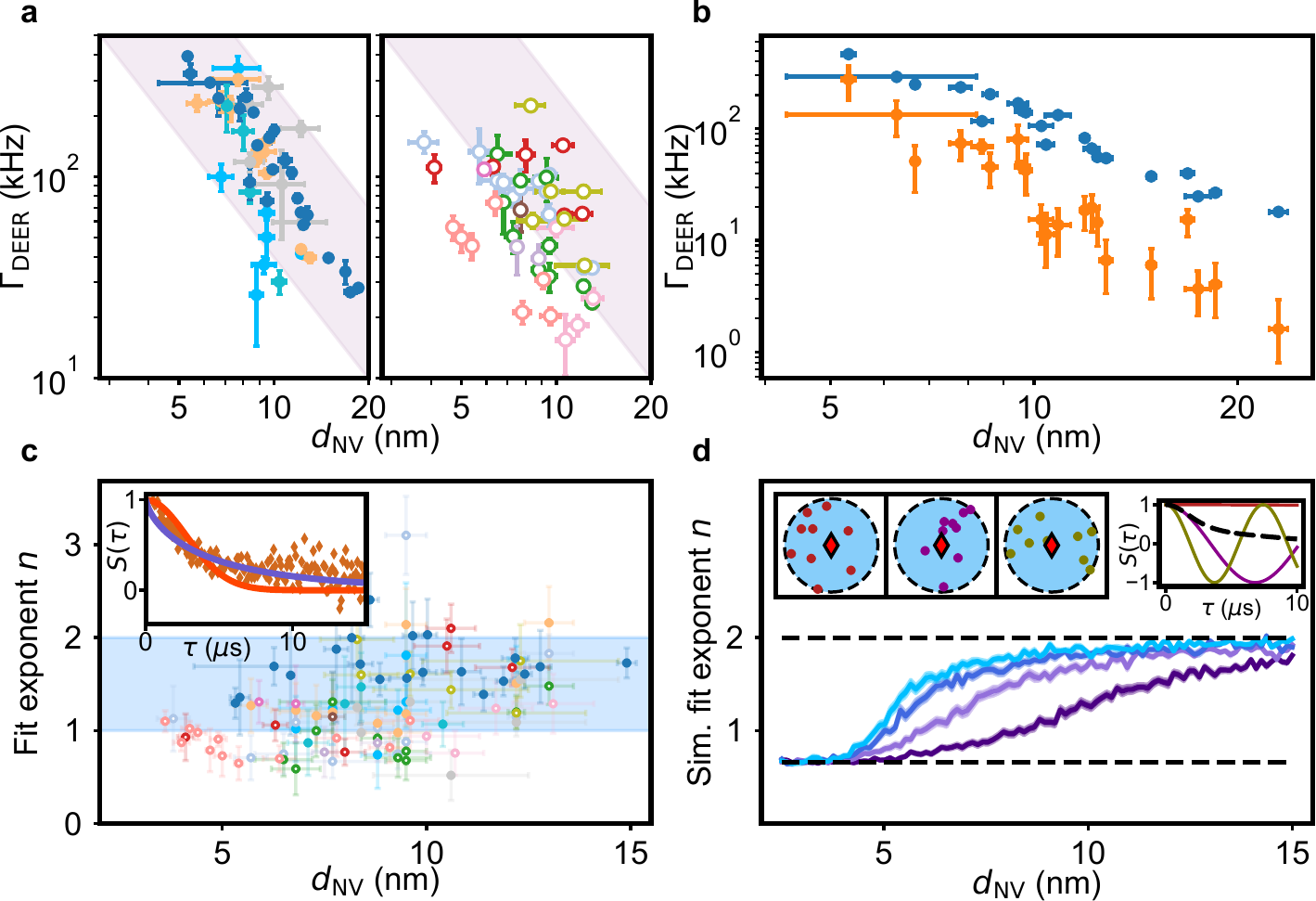}
 \caption{\textbf{DEER coupling rates and stretching factors.}
 \textbf{a}, Fitted DEER couplings as a function of depth for numerous NV centers. All samples in the left plot have nominally identical surface preparations (activated or reset surface, SI), while samples in the right plot represent different surface preparations described in the Supplementary Information. Shaded region indicates expected couplings for surface spin densities of 0.001-0.05 nm$^{-2}$, according to Eq.~\eqref{eq:coupling}. \textbf{b}, DEER coupling vs depth for the same NV centers in \bosample{} before (blue) and after (orange) annealing at 650$^\circ$ C. \textbf{c}, Fitted exponent as a function of depth for samples in \textbf{a}. Filled markers are samples with similar surfaces. Shaded region indicates exponents between 1 and 2. Inset: example DEER coupling curve from \elanasample{} (orange points) fit with a fixed $n = 2$ (red) and with $n$ as a free parameter (purple). $n = 0.7$ is a better fit. \textbf{d}, Fits of exponent vs depth for simulated data. Colors indicate different simulation spin densities of 0.008 nm$^{-2}$ (light blue), 0.006 nm$^{-2}$ (dark blue), 0.004 nm$^{-2}$ (light purple), and 0.002 nm$^{-2}$ (dark purple). Shaded regions are confidence intervals for fits of simulated data. Dashed lines indicate limiting values of $n = 2$ and $n = 2/3$. (Inset) Simulation of representative spin configurations (dots) relative to an NV center that is 5 nm deep (red diamond). Simulation radius is 25 nm. DEER signals for the representative configurations are shown in the last panel. Dashed line is an average of 10,000 iterations. All error bars are 68\% confidence intervals of the fits.}
 \label{fig:Fig2}
 \end{figure*}

We fit the resulting FID to an exponential decay of the form $S(\tau) = e^{-\left(\Gamma_{\rm{DEER}} \tau\right)^n}$, where $\tau$ is the total free evolution time, $\Gamma_{\rm{DEER}}$ is the DEER decay rate, and $n$ is the exponential stretching factor. The fitted decay rates for 105 NV centers in nine diamond samples are plotted in Fig.~\ref{fig:Fig2}a. For samples with similar processing histories (left plot), the couplings for a given depth cluster together, while samples with different surface preparations (right plot) show different offsets, which we interpret as arising from different densities ($\sigma$) of surface spins (\hyperref[sec:sampleprocessing]{Supplementary \ref{sec:sampleprocessing}}).

To demonstrate control over the surface spin density, we anneal one diamond sample at 650$^\circ$C in vacuum and measure the same NV centers before and after this procedure (\hyperref[sec:spinremoval]{Supplementary \ref{sec:spinremoval}}). In Fig.~\ref{fig:Fig2}b, we plot $\Gamma_{\rm{DEER}}$ for the same NV centers before and after this annealing procedure and find that, indeed, the DEER signal is significantly reduced following annealing, suggesting a reduction in the density of surface spins. We also characterize NV center coherence before and after annealing and observe that changes in coherence are minimal (\hyperref[sec:spinremoval]{Supplementary \ref{sec:spinremoval}}). We note that even carefully prepared diamond surfaces host numerous other electronic traps that can be inferred from X-ray spectroscopy methods \cite{sangtawesin2019origins}, and that in our samples these other traps are likely dominant sources of electric and magnetic field noise at the surface.

Previous work has treated the surface spin bath as a dense, 2D ensemble of spins with positions fixed but spin states that are initialized randomly between sequences \cite{Sushkov2014,Grinolds2014,Kim2015, bluvstein2019extending}. The DEER decay rate can then be computed as
\begin{equation}
    \Gamma_{\rm{DEER}} =\left(\frac{\mu_0}{4\pi}\right)\frac{\sqrt{3\pi\sigma}\gamma_e^2\hbar}{8 d_{\rm{NV}}^2}\,,
    \label{eq:coupling}
\end{equation}
 \noindent{}where $\mu_0$ is the permeability of free space, $\gamma_e$ is the electron gyromagnetic ratio in units of radians/second, and $\hbar$ is the reduced Planck's constant. This decay arises from an ensemble average over many realizations of spin state configurations, with fixed locations, in which there is no dominant proximal spin. The resulting distribution of total magnetic field amplitudes sensed by the NV center is then normally distributed, which is a requirement for the derivation of equation~\eqref{eq:coupling} and also leads to $n = 2$ in the exponential decay \cite{deSousa2009}(\hyperref[sec:derivation]{Supplementary \ref{sec:derivation}}).

 Using this model to extract surface spin densities, we find that much of our data fall well outside the regime where the assumptions of equation~\eqref{eq:coupling} are valid. The lower limit of the shaded region in Fig.~\ref{fig:Fig2}a is plotted from equation~\eqref{eq:coupling} for a surface spin density of 0.001 nm$^{-2}$, or an average nearest neighbor spin-spin separation of $\approx15$ nm (\hyperref[sec:derivation]{Supplementary \ref{sec:derivation}}), greater than the depth of the majority of NV centers measured here. For such a low density of spatially fixed surface spins, equation~\eqref{eq:coupling} should not be valid because the NV center primarily senses the field from the few nearest surface spins and hence the distribution of total magnetic field amplitudes is no longer Gaussian. Additionally, different NV centers should statistically sample different surface spin configurations, leading to a large variation in $\Gamma_{\rm{DEER}}$ among NV centers \cite{bluvstein2019extending}. Instead, we observe that $\Gamma_{\rm{DEER}}$ is fairly consistent at similar depths, and in particular does not exhibit increased variance for shallower depths, which is expected due to the increased likelihood of finding a few dark spins that are markedly closer to the NV center than the others. In the extremely dilute limit, the NV center would primarily sense the field from a single surface spin, which results in strong coherent oscillations rather than an exponential decay. While some of these strongly coupled surface spins have been observed previously \cite{Sushkov2014, Grinolds2014} and were found in some of our nine samples (\hyperref[sec:sampleprocessing]{Supplementary \ref{sec:sampleprocessing}}), these events are relatively rare,  occurring in  $< 2\%$ of measured NV centers.

  \begin{figure*}[hbt!]
  \includegraphics[width=6.75in]{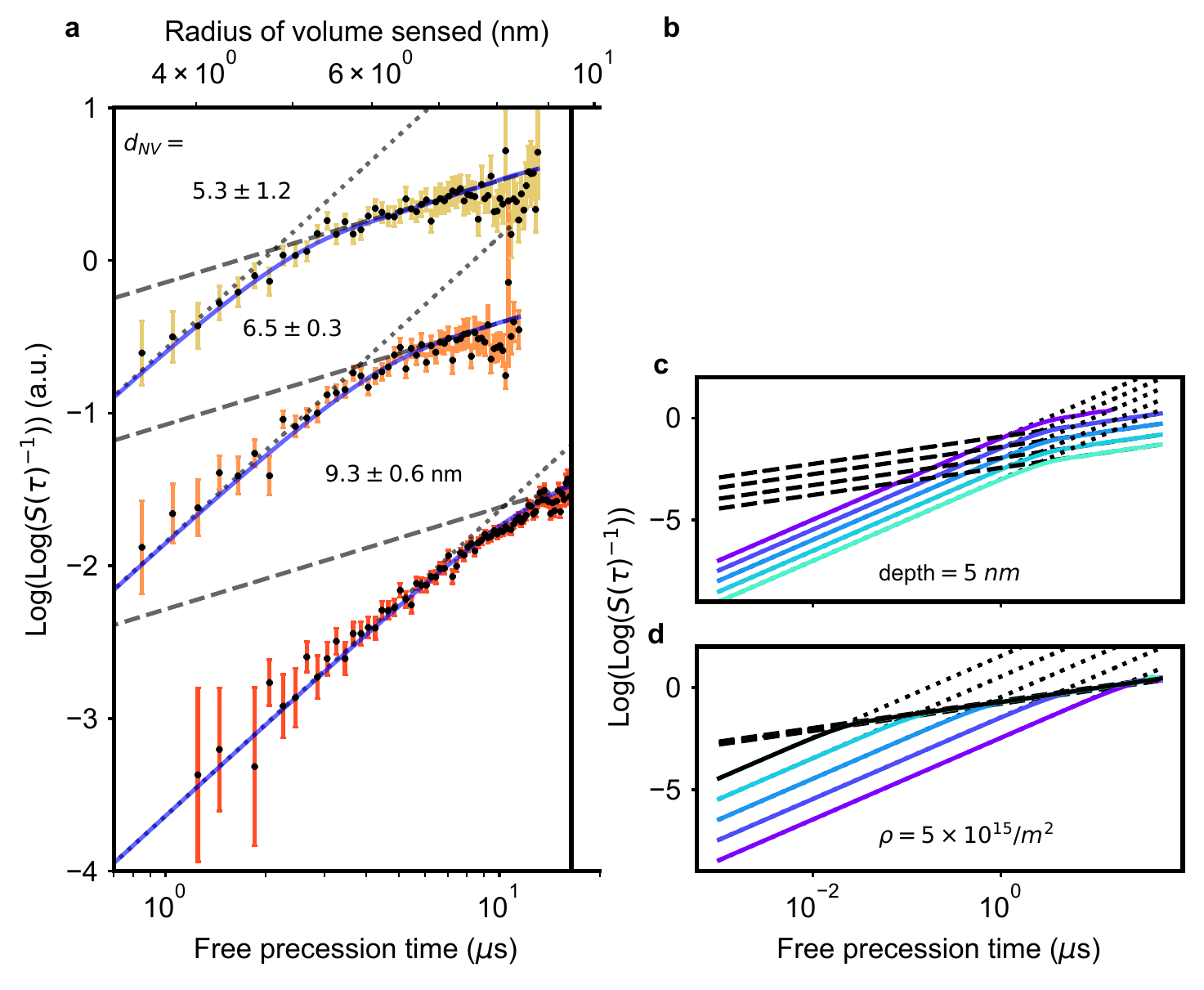}
  \caption{\textbf{Temporal transition of the stretching factor.} \textbf{a}, DEER echo signal from shallow NV centers in \elanasample{} with three different depths extracted from the proton NMR signal. Error bars are the standard error of the mean due to photon shot noise. The purple line is a fit with the analytical result for configurational averaging (SI). Dashed line is $n = 2$, dotted line is $n = \frac{2}{3}$. Curves are offset for clarity. \textbf{b}, Diagram depicting the expanding strong dipolar coupling volume of the NV center as the duration of the echo sequence increases. The form of the signal detected by the NV center changes as the volume intersects with the 2D plane of the surface spins. \textbf{c}, Simulations of the DEER echo signal detected from a 5 nm deep NV center near a surface with a logarithmically spaced density of 10$^{-4}$ nm$^{-2}$ (blue) to 10$^{-2}$ nm$^{-2}$ (purple) where surface spins undergo configurational averaging. The dashed lines correspond to $n = 2$, and the dotted lines correspond to $n = \frac{2}{3}$. \textbf{d}, Simulation of the DEER echo signal detected from NV centers with log spaced depths from 1 nm (blue) to 10 nm (purple) with configurational averaging of the surface spins. The density of simulated surface spins is 0.005 nm$^{-2}$. The solid black line is the analytic solution for a 1 nm deep NV center (SI).}
  \label{fig:Fig3}
  \end{figure*}

In Fig.~\ref{fig:Fig2}c, we examine the fitted exponent across all of the samples depicted in Fig.~\ref{fig:Fig2}a. For the majority of the shallow NV centers we investigated, a Gaussian decay profile provides a poor fit to the data, as is demonstrated in Fig.~\ref{fig:Fig2}c (inset). Furthermore, we observe that shallower NV centers exhibit lower fitted exponents (Pearson's correlation coefficient 0.4, p \textless 1.5$\times 10^{-5}$), and that many NV centers exhibit $n < 1$. The exponential stretching factor $n$ is related to the form of the noise generated by the environment spins \cite{deSousa2009}. In particular, a Gaussian ($n = 2$) decay occurs for a quasi-static bath, while a smaller exponent may arise from a finite bath correlation time. For a white noise bath, this exponent saturates at $n = 1$, so the observation of $n < 1$ for many NV centers is surprising. Additionally, measurements of the surface spin $T_1$ times for a select number of NV centers in this sample (\hyperref[sec:surfacespinmeasurements]{Supplementary \ref{sec:surfacespinmeasurements}}) find surface spin $T_1$ times are generally $>$30 $\mu$s, which is long compared to the spin echo decay time of the shallow NV centers, and so the quasi-static assumption is justified. We also note that although a short surface spin $T_1$ time can reduce the value of $n$ from 2 towards 1, the stretching factor as a function of depth should display a trend opposite to that of Fig.~\ref{fig:Fig2}c (\hyperref[sec:effectsoft1]{Supplementary \ref{sec:effectsoft1}}).

 Finally, for a few NV centers, we sample the coherence decay with much finer time steps to interrogate the shape of the decay in more detail. In Fig.~\ref{fig:Fig3}a, we plot the logarithm of the inverse DEER signal, $\log \left\{S(\tau)^{-1}\right\}$, as a function of free precession time on a log-log plot so that the stretching factor is given by the slope of the line. Instead of a single slope, we observe a transition in the stretching factor at some critical time, which increases with increasing NV center depth.

In order to explain our observations showing the lack of statistical variation among NV centers, the lack of coherent coupling to surface spins, the observed anomalously small decay exponents, and the transition in decay exponents with time, we consider a model wherein most of the surface spins do not have permanently fixed locations, but can occasionally ``hop'' between unoccupied sites such that the position is stationary for one experimental sequence and may change between sequences, as depicted in Fig.~\ref{fig:Fig1}a. The NV center then effectively senses a surface spin ensemble averaged over many positional configurations. For a dilute bath, such an averaging drastically changes the form of the decay due to strong fluctuations of the dipolar coupling strength stemming from the shot-to-shot fluctuations in the spin positions \cite{klauder1962spectral, dobrovitski2008decoherence}. In particular, the stretched exponents arise naturally from sampling different configurations of the spin bath, each with its own characteristic decay time \cite{dobrovitski2008decoherence, Stanwix2010Coherence}. Other electronic defects in diamond are known to change occupation under the intense illumination of the 532 nm laser used for NV center polarization and readout \cite{Siyushev2013Optically, wolters2013measurement}, and so it is conceivable that optical initialization and readout are responsible for inducing reconfiguration. The required averaging in our experiments would make it difficult to determine the frequency of reconfiguration or to observe an effect from illumination power, as, due to the large number of experimental repetitions typical of NV center experiments ($\sim 10^6$ repetitions), even a small probability of hopping results in sampling many spatial spin distributions.

In order to quantitatively investigate our model, we performed Monte Carlo simulations of the DEER signal resulting from configurational averaging of a low density of surface spins, as shown in Fig.~\ref{fig:Fig2}d (insets). By fitting the resulting FID of a large number (10$^5$) of averages to an exponential decay with a variable exponent $n$, the depth dependence of $n$ is readily apparent and is in qualitative agreement with the trend observed in Fig.~\ref{fig:Fig2}c. The fitted values of $n$ approach 2 for large depths ($d_{\rm{NV}} > 1/(2\sqrt{\sigma})$), but gradually transition to a value of 2/3 as the depth is decreased below the average surface spin separation.

By integrating over the possible configurations of surface spins \cite{lacelle1993typical}, we can obtain an analytic expression for the DEER signal measured by an NV center external to the spin bath (\hyperref[sec:derivation]{Supplementary \ref{sec:derivation}}). For short times such that $\frac{\mu_0}{4\pi}\gamma_e^2\hbar \tau \ll d_{\rm{NV}}^3$, we recover equation~\eqref{eq:coupling} and $n=2$, even though the spins are assumed to be non-stationary in this case. In the opposite limit, $\frac{\mu_0}{4\pi}\gamma_e^2\hbar \tau \gg d_{\rm{NV}}^3$, we find instead

\begin{equation}
    \langle S(\tau)\rangle_c=\exp\left(-\frac{9\sqrt{\pi}\Gamma(\frac{11}{6})\sigma}{5}\left(\frac{\mu_0\gamma^2\hbar}{8\pi}\right)^{2/3}\tau^{2/3}\right)\,,
    \label{eq:hopping_longtime}
\end{equation}

\noindent{}where $\Gamma(x)$ is the Gamma function. The stretching factor $n=2/3$ for a 2D bath of spins has been predicted previously \cite{fel1996configurational, hahn2019long}, and its appearance here implies that for long times the entire system, including the sensor, behaves as if it is two dimensional.

In our finely sampled data in Fig.~\ref{fig:Fig3}a, we find that the data are well-fit by our analytical model, which captures the transition regions that move to later times for increasing depths. Because the location of this transition is sensitive to depth alone, it is possible to extract a depth for each NV center from these measurements independent of the surface spin density. These extracted depths are in good agreement with those measured by proton NMR, which additionally indicates that the spins we address are indeed at the diamond surface (\hyperref[fig:additionaldata]{Supplementary Figure \ref{fig:additionaldata}}). Further model comparison and additional datasets can be found in the Supplementary Information (\hyperref[sec:addtionalmodelanalysis]{Supplementary \ref{sec:additionalNVs}-\ref{sec:addtionalmodelanalysis}}).

\begin{figure*}[hbt!]
\includegraphics[width=6.75in]{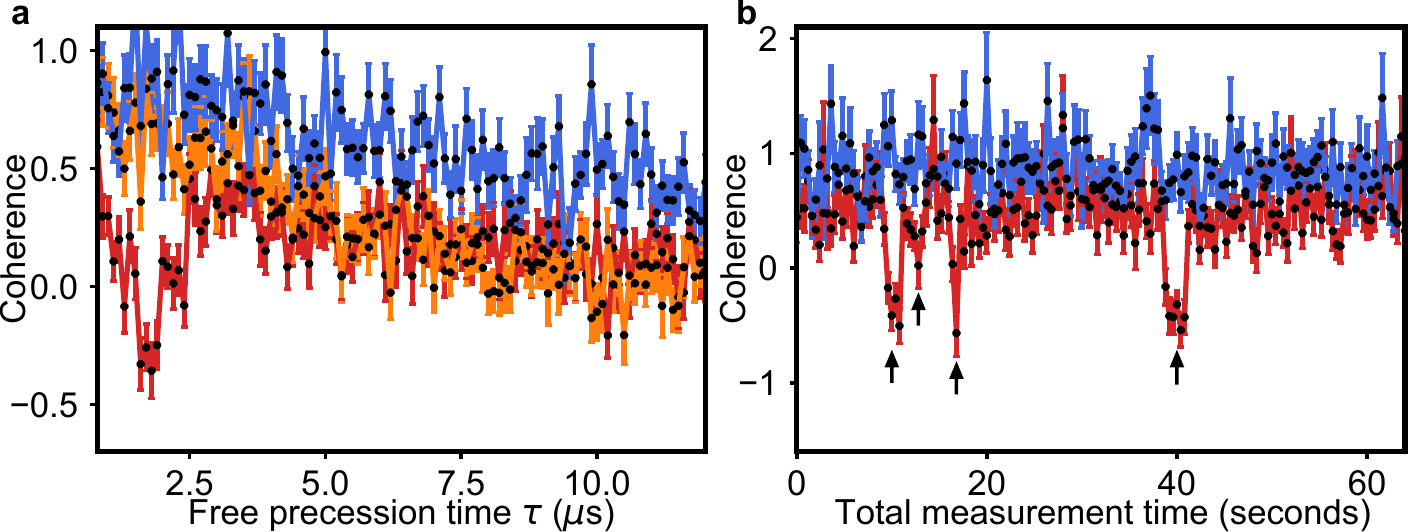}
\caption{\textbf{Transient strong coupling of surface spins.} \textbf{a}, Coherence (blue), coherently coupled DEER signal (red), and incoherent bath signal (orange) from the same NV center measured at different times. \textbf{b}, Time series for NV center in \textbf{a}, with a fixed free precession interval of 2 $\mu$s. The DEER signal drops below 0 for short intervals, indicating transient coherent coupling to a nearby dark spin (black arrows). Error bars are the standard error of the mean due to photon shot noise.}
\label{fig:Fig4}
\end{figure*}

In Fig.~\ref{fig:Fig3}c and Fig.~\ref{fig:Fig3}d, we examine the simulated behavior of $n$ in more detail by plotting $\log \left\{S(\tau)^{-1}\right\}$. We note a change in the stretching factor from $n = 2$ to $n = 2/3$ at a depth-dependent critical time, similar to what is observed in the data in Fig.~\ref{fig:Fig3}a. The shift of this transition to earlier times for shallower NV centers indicates that the intermediate values of $n$ measured in Fig.~\ref{fig:Fig2}c and simulated in Fig.~\ref{fig:Fig2}d are not the result of a continuously varying exponent, but rather of averaging different numbers of points that lie in either the $n = 2$ or $n = 2/3$ regions of the decay and fitting to a single exponent.

In some cases, re-positioning of surface spins can be detected more directly. In Fig.~\ref{fig:Fig4}a, we show DEER traces from the same NV center measured minutes apart, showing both a coherent coupling signal that dips below 0 (red), and a weak incoherent coupling signal observed a short time later (orange). By binning into smaller numbers of averages for a fixed free precession interval of 2 $\mu$s, it is possible to see the switching behavior in real time in Fig.~\ref{fig:Fig4}b. We interpret these results as indicating the presence or absence of a particularly stable proximal spin, and hence as evidence that the occupation of such traps can change. While our model does not rule out the existence of some number of stable, spatially fixed spins, based on the percentage of measured NV centers in our samples that display coherent oscillations, we estimate an upper bound on the number of stationary surface spins of $\approx 10\%$ of the total spins detectable by DEER (\hyperref[sec:fractionhopping]{Supplementary \ref{sec:fractionhopping}}).

The time-dependent behavior of the exponent $n$ can be understood by considering the dipolar coupling between an NV center and the sheet of surface spins at a distance $d_{\rm{NV}}$ away. Initially, the NV center senses a small contribution from all of the spins, which results in a Gaussian ($n = 2$) decay. Once the phase accumulated from the nearest spin reaches a large enough value ($\frac{\mu_0}{4\pi}\frac{\gamma_e^2\hbar}{d_{\rm{NV}}^3}\tau \sim 2\pi$), further time evolution contributes only an oscillatory phase while more distant spins still contribute a summation of many small phases. This process can be considered with a simple geometric picture. The quantity $(\frac{1}{2\pi}\frac{\mu_0}{4\pi}\gamma_e^2\hbar\tau)^{1/3} \approx (52$ $\rm{MHz\cdot nm^3}\tau)^{1/3}$ defines the radius of a sphere, centered on the NV center, that increases with the free evolution time. Spins within this sphere contribute significant phase ($ \gtrsim \pi$) to the NV center. Initially, there are no surface spins within this sphere. The sphere of strong coupling will increase in size until it intersects the surface when $(52$ $\rm{MHz}\cdot\rm{nm}^3\tau)^{1/3} \approx d_{\rm{NV}}$, at which time spins begin to interact strongly with the NV center. As is depicted in Fig.~\ref{fig:Fig3}b, the number of strongly interacting spins grows with the cross-sectional area that intersects the surface plane, which grows like $\tau^{2/3}$. It is only once the volume of strong interactions grows to include the plane in which the spins dwell that the dimensionality of the bath impacts the NV center coherence, and the NV center decay reflects this dimensionality for long times. In a sense the NV center becomes a member of the 2D bath for sufficiently long interaction times, which is supported by the absence of any dependence on $d_{\rm{NV}}$ in the long-time limit given by equation~\eqref{eq:hopping_longtime}.

The inferred dynamics of the surface spins has immediate consequences for surface-spin-based sensing of external targets and for using surface spins as a model system to explore many-body interactions in two dimensions. In particular, the stability of these spins needs to be taken into account and may need to be enhanced. Samples prepared with different processing might host more stable surface spins \cite{Sushkov2014}, and it is possible that diamond surface treatments could be used to stabilize, rather than eliminate, these defects. Lacking this, other, potentially more stable systems, such as spinful defects in 2D materials \cite{liu2009photochemical, yazyev2007defect}, could be transferred onto the diamond surface and interrogated by NV centers. A more accurate determination of reconfiguration probability and mechanism could be made with single-shot readout of the NV center through resonant excitation \cite{robledo2011high, irber2020robust} or mapping the electronic spin state to the host nitrogen spin \cite{lovchinsky2016nuclear}. Furthermore, techniques such as nano-MRI \cite{Grinolds2014, grinolds2011quantum} or MRFM \cite{Rugar2004} could enable a more systematic survey of surface spin properties. Dilute spin ensembles in other materials can be studied with scanning NV center experiments \cite{maletinsky2012robust, Pelliccione2016b} where the distance to the sample can be tuned continuously. More generally, the techniques for understanding surface environment dynamics presented here may be broadly applicable in other quantum systems. Specifically, the dynamics of proximal probes can carry significant information: the minimum impurity separation, the dimensionality of the environment, and the scales of finite dimensions are all reflected in the behavior of the decay, constituting a promising tool for studying the disordered environment of quantum systems.

\newpage
\vspace{0.1in}
\noindent{}\emph{Note added:} During the completion of this work, we became aware of complementary work \cite{davis} using probe spins to measure the many-body noise from a strongly-interacting dipolar system, which will appear in the same arXiv posting. Both works use the decoherence profile of the probe spin to characterize the dimensionality and dynamics of the many-body system.

\section*{Acknowledgements}
	\noindent{}We thank Ania Jayich, Norman Yao, Soonwon Choi, Alastair Stacey, Cherno Jaye, and Andrew Evans for fruitful discussions. Undergraduate researchers Trisha Madhavan and Rohith Karur contributed to surface spin measurements that helped us to calibrate the annealing procedure. This work was supported in part by the DARPA DRINQS program (grants D18AC00015 and D18AC00033), the Center for Ultracold Atoms (NSF PHY-1734011), and the Moore Foundation (grant 4342.01). Sample surface preparation and spectroscopy was supported partially by the NSF under the CAREER program (grant DMR-1752047) and through the PCCM (grant DMR-1420541). LVHR acknowledges support from the DOD through the NDSEG Fellowship Program. EKU acknowledges support from the NSFGRFP (grant DGE1144152). DB acknowledges support from the NSFGRFP (grant DGE1745303) and the Hertz Foundation. SS acknowledges support from the PMU-B (grant B05F630108). JC acknowledges the financial support from the MST, Taiwan (MOST-109-2112-M-018-008-MY3). MF was supported by an appointment to the IC Postdoc Research Program by ORISE through an inter-agency agreement between the U.S. DOE and the ODNI. AG acknowledges the support from the NKFIH in Hungary for the Quantum Technology Program (grant 2017-1.2.1-NKP-2017-00001), the National Excellence Program (grant KKP129866), and the EU QuantERA project (grant NN127889) and from the European Commission for the ASTERIQS project (Grant No. 820394). VVD was partially supported by QuTech Physics Funding (QTECH, program 172, No. 16QTECH02) which is partly financed by the NWO. This work was performed in part at the Imaging and Analysis Center at Princeton, the Center for Nanoscale Systems at Harvard (NSF Grant 1541959), and the research beam line U7A of the National Synchrotron Light Source operated for the DOE by Brookhaven National Laboratory (Contract DE-AC02-98CH10886).

\section*{Contributions}
	\noindent{}NPdL, BLD, and KDG conceived of the experiment. BLD, EU, LVHR, SS, YN, MF, ZY, and ELP performed the experiments and data analysis. DB, HZ, JC, AG, and VVD provided theoretical support and interpretation of results. BLD, LVHR, EKU, DB, SS, and NPdL wrote the manuscript and prepared figures with input from the other authors. MDL and NPdL supervised the work.

\section*{Supplementary Information}
  \noindent{}Supplementary information is available for this paper.

\section*{Correspondence}
  \noindent{}Correspondence and requests for materials should be addressed to Nathalie P. de Leon npdeleon@princeton.edu.

\newpage

\renewcommand{\thefigure}{\textbf{S}\arabic{figure}}
\renewcommand{\theequation}{S\arabic{equation}}
\renewcommand{\thetable}{\rm{S}\arabic{table}}
\setcounter{figure}{0}
\setcounter{equation}{0}

\renewcommand\thesection{\textbf{S}\textbf{\arabic{section}}}
\renewcommand\thesubsection{\thesection.\arabic{subsection}}

\onecolumngrid
\begin{center}
\textbf{ \Large Supplementary Information for \papertitle}
\end{center}
\tableofcontents
\newpage

\section{Measurement setup}
\label{sec:setup}
The details of our measurement setup have been published elsewhere \cite{sangtawesin2019origins}, but we briefly review them here for completeness. The data for this manuscript were taken on several measurement setups that were constructed in a similar manner but with different components.

NV center measurements were performed on a home-built confocal microscope. NV centers are excited by a 532 nm frequency doubled Nd:YAG solid state laser which is modulated with an acousto-optic modulator. The beam is scanned using galvo mirrors and projected into an oil immersion objective (Nikon, Plan Fluor 100X,NA = 1.30) with a telescope in a 4f configuration.  Laser power at the back of the objective was kept between 60–100 $\mu$W, approximately 25\% of the saturation power of a single NV center, in order to avoid irreversible photobleaching. A dichroic beamsplitter separates the excitation and collection pathways, and fluorescence is measured using a fiber-coupled avalanche photodiode (Excelitas SPCM-AQRH-44-FC). A neodymium permanent magnet was used to introduce a DC magnetic field for Zeeman splitting, and the orientation of the magnetic field was aligned to within 1$^\circ$ of the NV center axis using a goniometer. Measurements were performed at moderate magnetic fields of $\approx$ 300 Gauss so that the surface spin and NV center transitions were frequency-resolved.

Spin manipulation on the NV center and surface spins was accomplished using microwaves. Two signal generators, which output the NV center and DEER transition frequencies, were separately gated with fast SPDT switches (Mini-Circuits ZASWA-2-50DR+). In one set up, these signals were then amplified by Mini-Circuits ZHL-16W-43+ and ZHL-100W-13+, respectively before being combined with a high-power resistive combiner and delivered to the sample via a coplanar stripline. In another set up, the gated signals were combined with a resistive combiner (Mini-Circuits ZFRSC-42-S+) and amplified with a high-power amplifier (Ophir 5022A) before being delivered to the sample via a coplanar stripline. The stripline was fabricated by depositing 10 nm Ti, 1000 nm Cu, and 200 nm Au on a microscope coverslip. Following metallization, the stripline was photolithographically defined and etched with gold etchant and hydrofluoric acid.  Finally, a 100 nm layer of Al$_2$O$_3$ was deposited on top of the fabricated stripline via atomic layer deposition (ALD) to protect the metal layer. Pulse timing was controlled with a Spincore PulseBlaster ESR-PRO500 with 2 ns timing resolution. The DEER pulse was either applied simultaneously to the NV center pulse, as depicted in the Main Text Fig.~1, or it was offset from the NV center pulse in order to avoid saturation of the amplifier in some cases. In the latter case, an identical DEER pulse was applied immediately after the first $\pi/2$ pulse on the NV center as in \cite{Mamin2012}.

\section{Sample processing}\label{sec:sampleprocessing}
All of the samples measured for this work were electronic grade samples ($<$5 ppb nitrogen, $<$1 ppb boron) from Element 6. Those with an additional overgrown $^{12}$C enriched layer are indicated in Table \ref{tab:sampletable}. Most of the samples had surfaces that were prepared prior to implantation following the procedure outlined in \cite{sangtawesin2019origins}. Briefly, the pre-implantation surface preparation begins with scaife polishing, followed by etching in Ar/Cl$_2$ followed by O$_2$ plasma, and then annealing to 1200$^{\circ}$C in a vacuum tube furnace to graphitize and remove the first few nanometers of the surface. \NF{} and \elanasample{} did not undergo this surface preparation prior to implantation and were implanted with the ``as grown'' surface. All samples were implanted with $^{15}$N, and annealed to 800$^{\circ}$C in a tube-vacuum furnace to form NV centers.

After the activation anneal, NV centers were measured under a variety of different surfaces, as noted in Table \ref{tab:sampletable}. In some cases, DEER coupling was characterized in the same sample after different surface processing steps. The final surface conditions noted in Table \ref{tab:sampletable} were prepared as follows:
\begin{itemize}

    \item 	Activated: NV centers were measured directly after 800$^{\circ}$C activation anneal and triacid cleaning (reflux sample in 1:1:1 sulfuric, nitric, and perchloric acids for 2 hrs)
    \item 	Oxygen annealed: NV centers were measured after heating the sample to 440-460$^{\circ}$C while flowing an atmosphere of oxygen gas over the samples, followed by cleaning in a 2:1 piranha mixture (sulfuric acid and hydrogen peroxide) for 20 min.
    \item Reset: NV centers were measured after `resetting' the surface to a state comparable to the activated surface by reannealing the sample to 800$^{\circ}$C in a vacuum tube furnace at around 10$^{-6}$ mbar, and subsequently triacid cleaning the sample.
    \item 500$^{\circ}$C annealed: NV centers were measured after annealing the sample to 500$^{\circ}$C in a vacuum tube furnace, followed by triacid cleaning.

\end{itemize}

\begin{table}[H]
\begin{center}
\begin{tabular}{ |c|c|c|c|c| }
\hline
Sample & C12 enriched & Pre implant surface prep & Color & Surface condition \\
\hline\hline
 \multirow{4}{*}{\GA} & \multirow{4}{*}{}  & \multirow{4}{*}{$\times$}&\cellcolor{GA_0} & activated \\
\cline{4-5}
 & &  & \cellcolor{GA_1_O2} & oxygen annealed \\
\cline{4-5}
 & & &\cellcolor{GA_1_Reset} & reset \\
\cline{4-5}
 & & &\cellcolor{GA_2_O2} & oxygen annealed \\
\hline
\CT & & $\times$ & \cellcolor{CT_1200}& oxygen annealed \\
\hline
\NF & $\times$ & &\cellcolor{NF_0} & activated \\
\hline
\PA & & $\times$ &\cellcolor{PA_3_Reset} & reset \\
\hline
\multirow{2}{*}{\MB} & \multirow{2}{*}{$\times$} & \multirow{2}{*}{$\times$} & \cellcolor{MB_1_O2}& oxygen annealed \\
\cline{4-5}
 & &  &\cellcolor{MB_2_Reset} & reset \\
\hline
\multirow{3}{*}{\NJ} & \multirow{2}{*}{}& \multirow{2}{*}{$\times$} &  \cellcolor{NJ_2_O2} & oxygen annealed \\
\cline{4-5}
 & &   &\cellcolor{NJ_2_Reset} & reset \\
\cline{4-5}
 & & &\cellcolor{NJ_3_500C} & 500$^{\circ}$C annealed \\
\hline
\Toluca & & $\times$ & \cellcolor{Toluca_0}& activated \\
\hline
\bosample & $\times$ & $\times$ & \cellcolor{bosample}& activated \\
\hline
\elanasample & $\times$& &\cellcolor{elanasample} & oxygen annealed\\

 \hline

\end{tabular}
\caption{Diamond samples \label{tab:sampletable}}
\end{center}
\end{table}

\section{Fitting procedures}
\label{sec:fittingprocedures}
DEER datasets consist of a bright (m$_{\rm{s}}$ = 0 projection) and dark (m$_{\rm{s}}$ = -1 projection) reference (BrightRef and DarkRef, respectively), spin echo signals projected onto m$_{\rm{s}}$ = 0 and m$_{\rm{s}}$ = -1 (sig$_0$ and sig$_1$, respectively), and DEER projected onto m$_{\rm{s}}$ = 0 and m$_{\rm{s}}$ = -1 (deer0 and deer1, respectively). There are also error bars for each point in each of these categories, given by the variance of the counts, which is $\sqrt{N_{\rm{photons}}}$ for the shot-noise limited NV center spin readout employed here. These signals are then averaged to calculate coherence decay points, $C_i$:

\begin{equation}
C_i = \left(\frac{\rm{Sig}_{0i} - \rm{Sig}_{1i}}{\rm{BrightRef}_i - \rm{DarkRef}_i}\right)
\end{equation}

\noindent and DEER measurement points, $D_i$, in the same fashion. The error bars for the coherence are then given by

\begin{equation}
\sigma_{Ci}^2 = \left(\frac{1}{\rm{BrightRef}_i - \rm{DarkRef}_i}\right)^2(\sigma_{0i}^2 + \sigma_{1i}^2) + \left(\frac{\rm{Sig}_{0i} - \rm{Sig}_{1i}}{(\rm{BrightRef}_i - \rm{DarkRef}_i)^2}\right)^2(\sigma_{Bi}^2 + \sigma_{Di}^2)
\end{equation}

\noindent An equivalent expression can be written down for the DEER signal error bars.

In order to remove the effects of other decoherence mechanisms from the DEER measurement and obtain a free induction decay curve due to the surface spins alone, we divide the DEER signal by the coherence:

\begin{equation}
    S_i = \frac{D_i}{C_i}
\end{equation}

\noindent and propagate the error bars

\begin{equation}
\sigma_{Si}^2 = \left(\frac{\sigma_{Di}}{C_i}\right)^2 + \left(\frac{\sigma_{Ci}D_i}{(C_i)^2}\right)^2
\end{equation}

For the data presented in Fig.~1, we then fit to the stretched exponential function

\begin{equation}
f(\tau; a, \Gamma_{\rm{DEER}}, n) = a e^{-(\Gamma_{\rm{DEER}}\tau)^n}
\end{equation}

\noindent in order to obtain DEER decay rates ($\Gamma_{\rm{DEER}}$ and stretching factor $n$). The parameter $a$ is experimentally found to be very close to 1.

The data presented in Fig.~3 undergo further processing to better observe the transition in $n$ with time. Taking the logarithm of $1/f(\tau)$ and plotting on a log-log scale gives a line with a slope given by $n$ and an intercept of $n\log_{10}\Gamma_{\rm{DEER}}$.

First a conservative cutoff of $D_i > 0.1$ is introduced to remove sections of data with low signal to noise and to avoid infinities in taking the logarithm. We then calculate

\begin{equation}
    g_i = \log_{10}\ln\frac{1}{S_i}
\end{equation}

\noindent and the error bars of $g_i$
\begin{equation}
\sigma_{g_i} = \frac{\sigma_{Si}}{\ln(10) S_i\ln(S_i)}
\end{equation}

For many of the NV centers presented in Fig.~2, the DEER coupling rates are extremely small and in many cases consistent with 0. The processing and fitting procedure defined above was found to give poor results and instead a Markov Chain Monte Carlo procedure was employed to simultaneously fit the coherence and DEER data without division. The coherence was assumed to follow a function of the form

\begin{equation}
    C(\tau) = ae^{-(\tau/T_2)^c}
\end{equation}

While the DEER signal was assumed to follow

\begin{equation}
    D(\tau) = e^{-(\Gamma_{\rm{DEER}}\tau)^n}\cdot C(\tau)
\end{equation}

The data $C_i$ and $D_i$ were used to calculate likelihood functions for $C(\tau)$ and $D(\tau)$, respectively.

\section{Removal of surface spins through annealing}
\label{sec:spinremoval}

\subsection{X-ray spectroscopy}
\label{sec:xrayspec}
\begin{figure}[H]
    \centering
    \includegraphics[width=5in]{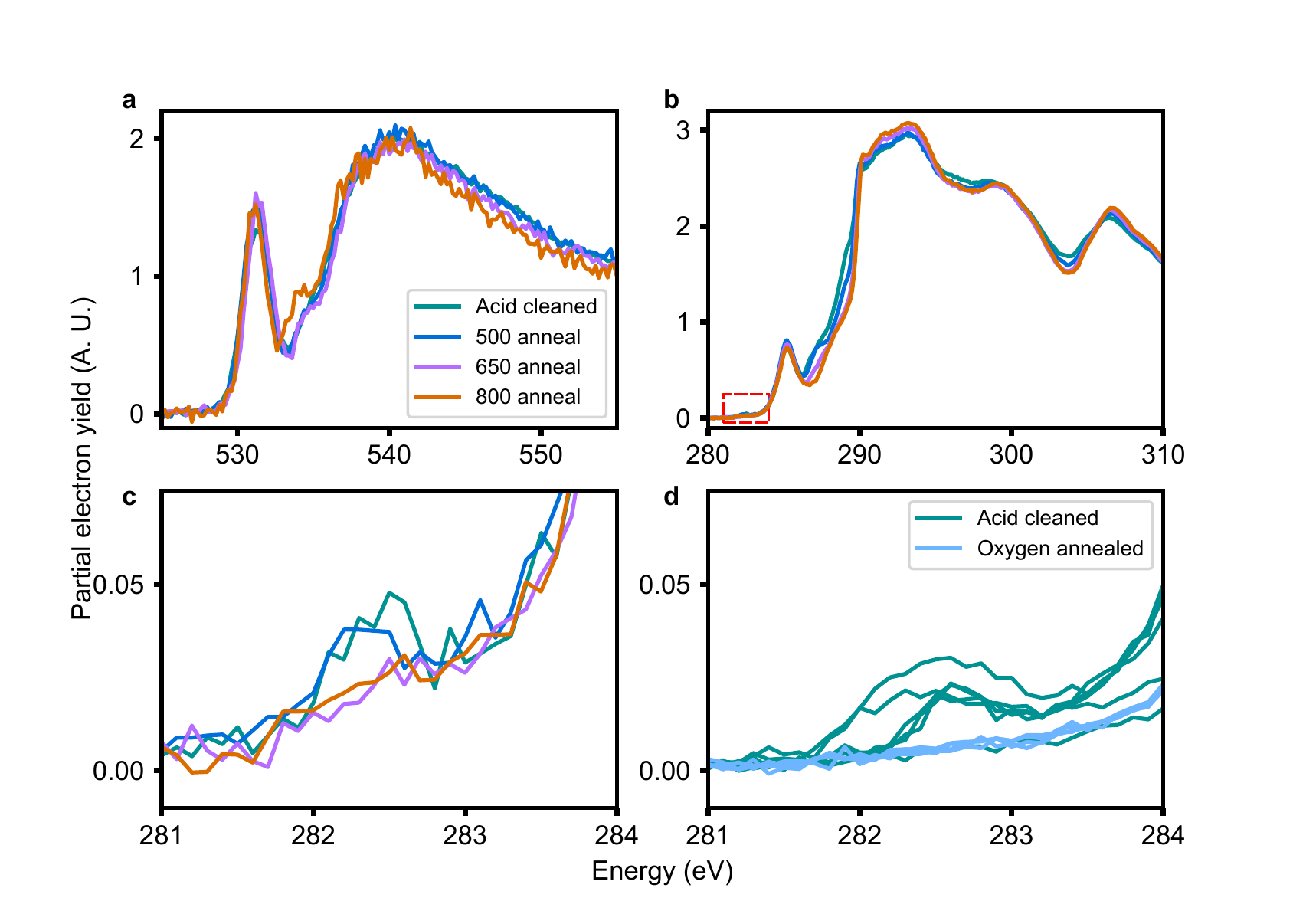}
    \caption{\textbf{NEXAFS through annealing.} \textbf{a}, Oxygen edge NEXAFS for diamond sample with initial acid cleaned surface (green), after 500$^\circ$C annealing (blue), 650$^\circ$C annealing (purple) and 800$^\circ$C annealing (orange). \textbf{b}, Carbon edge NEXAFS of the same annealing sequence on the same sample. Red box indicates location of dangling bond peak, shown in detail in \textbf{c}. \textbf{c}, Detailed view of dangling bond peak showing removal at higher annealing temperatures. \textbf{d}, Detailed view of dangling bond peak for several samples with initially prepared acid cleaned surfaces (green) and oxygen annealed surfaces (blue), showing absence of the peak for oxygen annealed surfaces. Oxygen annealing procedure and suppression of DEER signal are described in \cite{sangtawesin2019origins}.}
    \label{fig:NEXAFS}
\end{figure}

To investigate the microscopic origin of surface spins and to develop a procedure for their selective removal, we interrogated diamond surfaces using X-ray spectroscopy. In near-edge X-ray absorption fine structure (NEXAFS) spectroscopy, monochromatic incident X-rays excite core electrons, and the electron yield is measured as a function of X-ray energy, giving a signal that is proportional to the unoccupied density of states near the surface \cite{stohr1992springer}. We performed NEXAFS spectroscopy on diamond samples prepared with the standard surface preparation procedure described in section \ref{sec:sampleprocessing}, without implanting NV centers. NEXAFS spectroscopy at the carbon edge reveals a peak at 282.5 eV (Fig.~\ref{fig:NEXAFS}b). Previous studies assign this peak to sp$^3$ dangling bonds at the diamond surface \cite{bobrov2001electronic}. The presence of this dangling bond peak in the NEXAFS spectra correlates with the surface spin DEER signal measured via shallow NV centers. Specifically, we observe that the peak at 282.5 eV is present in thermally annealed and acid cleaned samples, while it is absent in samples after oxygen annealing as described in \cite{sangtawesin2019origins}. Quantitatively, oxygen annealing diminishes this peak from a signal to noise ratio of around four to a value below the noise floor (Fig.~\ref{fig:NEXAFS}(d)). Correspondingly, we have also previously shown that individual NV centers show decreased DEER coupling rates after oxygen annealing, and increased coupling after a thermal surface reset process \cite{sangtawesin2019origins}.

The correlation between the 282.5 eV peak in NEXAFS and DEER coupling strengths across samples suggests that dangling bonds at the diamond surface may be responsible for the surface spin signal detected in NV center-based sensing experiments. However, there are a variety of observable changes at the diamond surface upon oxygen annealing \cite{sangtawesin2019origins}, and a definitive link requires a process for selectively removing dangling bonds. When the samples are annealed in high vacuum at 650$^{\circ}$C for one hour, the dangling bond peak falls below the noise floor (Fig.~\ref{fig:NEXAFS}(c)), but the oxygen edge NEXAFS spectrum is essentially unchanged (Fig.~\ref{fig:NEXAFS}(a)). Thus, annealing at 650$^{\circ}$C leads to selective removal of the dangling bond peak. This annealing procedure also leads to a decrease in DEER coupling for the same NV centers, as shown in the Main Text.

\subsection{Annealing procedure}
\bosample{} was annealed in a vacuum tube furnace with a base pressure around 10$^{-6}$ mbar. The sample was heated from room temperature to 100$^{\circ}$C over 1 hour, held at 100$^{\circ}$C for 11 hours, heated to 650$^{\circ}$C over 20 hours, held at 650$^{\circ}$C for 2 hours, then cooled to room temperature.

We note that we have tried annealing in different vacuum systems and have sometimes measured a reduction in DEER coupling at inconsistent temperatures. We attribute these discrepancies to differences in annealing duration, temperature calibration, or thermal anchoring of sample to heater. Other samples showed a persistent DEER signal that was not reduced by vacuum thermal annealing or oxygen annealing. Many of these samples either did not go through processing to reduce subsurface damage prior to implantation or had rough surface morphologies from iterative processing.

\subsection{Coherence change with annealing}

\begin{figure}[H]
    \centering
    \includegraphics[width=2.5in]{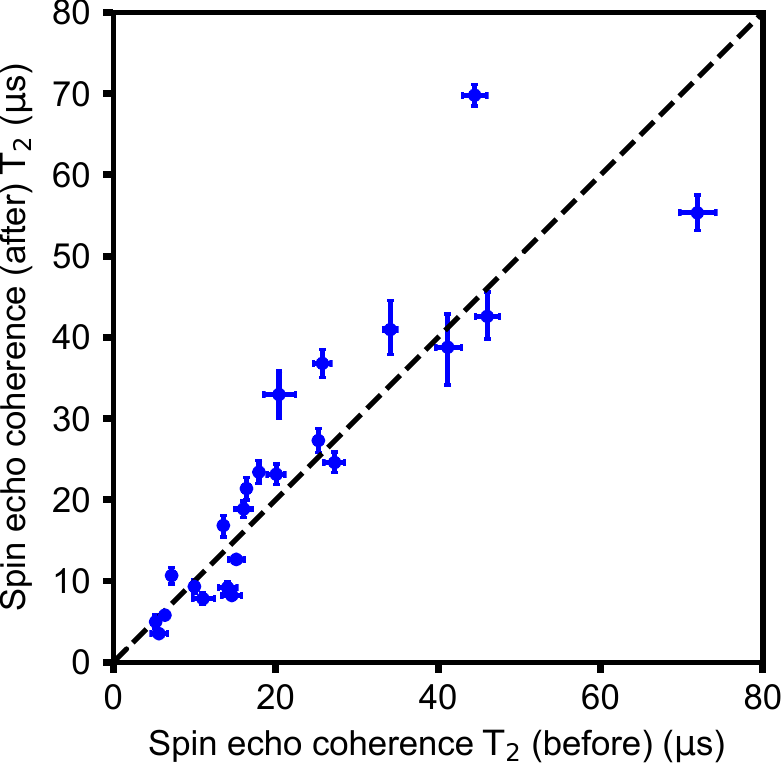}
    \caption{Coherence times of NV centers in \bosample{} after annealing, plotted as a function of pre-annealing coherence times. Dashed line has a slope of one.}
    \label{fig:T2beforeafter}
\end{figure}

For the NV centers depicted in Main Text figure Fig.~2a, we also measured the spin echo coherence time ($T_2$) of each NV center before and after annealing. The results are plotted in Fig.~\ref{fig:T2beforeafter}. In contrast to oxygen annealing coherence times of these NV centers are not significantly affected by the vacuum annealing procedure \cite{sangtawesin2019origins}. Because the DEER couplings for these NV centers are reduced by annealing, we interpret these results together to imply that the surface spin bath is not primarily responsible for limiting the coherence time of near surface NV centers in this sample.

\subsection{Spectral decomposition through annealing}
The spectral density of the noise bath, $S(\omega)$ can be probed by dynamical decoupling \cite{sangtawesin2019origins, Myers2014, Romach2015}. We use the non-symmetric XY8 pulse sequence \cite{wang2012comparison} with multiple repetitions to probe different regions of the noise spectrum.

In general, the coherence decay, $C(T)$, of an NV center is given by
\begin{equation}
\label{eq:coherence}
    C(T) = \exp(-\chi(T)),
\end{equation}
\noindent where T is the total free precession time ($T = N\tau$ for $N$ precession intervals of length $\tau$) and $\chi(T)$ is given by
\begin{equation}
\label{eq:spec}
    \chi(T) = \frac{1}{\pi}\int_0^\infty S(\omega)\frac{F_N(\omega T)}{\omega^2}d\omega
\end{equation}
\noindent where  $S(\omega)$ is the magnetic noise spectrum and $F_N(\omega T)$ is the N-pulse filter function given by
\begin{equation}
\label{eq:filter}
    F_N(\omega T) = 8\sin^4\left(\frac{\omega T}{4N}\right)\frac{\sin^2\left(\frac{\omega T}{2}\right)}{\cos^2\left(\frac{\omega T}{2N}\right)}.
\end{equation}

When $F_N(\omega T)$ is sharply peaked, Eq.~\eqref{eq:spec} can be approximated as

\begin{equation}
    \chi(T) \approx \frac{T S(\omega)}{\pi}.
\end{equation}

\begin{figure}[H]
    \centering
    \includegraphics[width=5in]{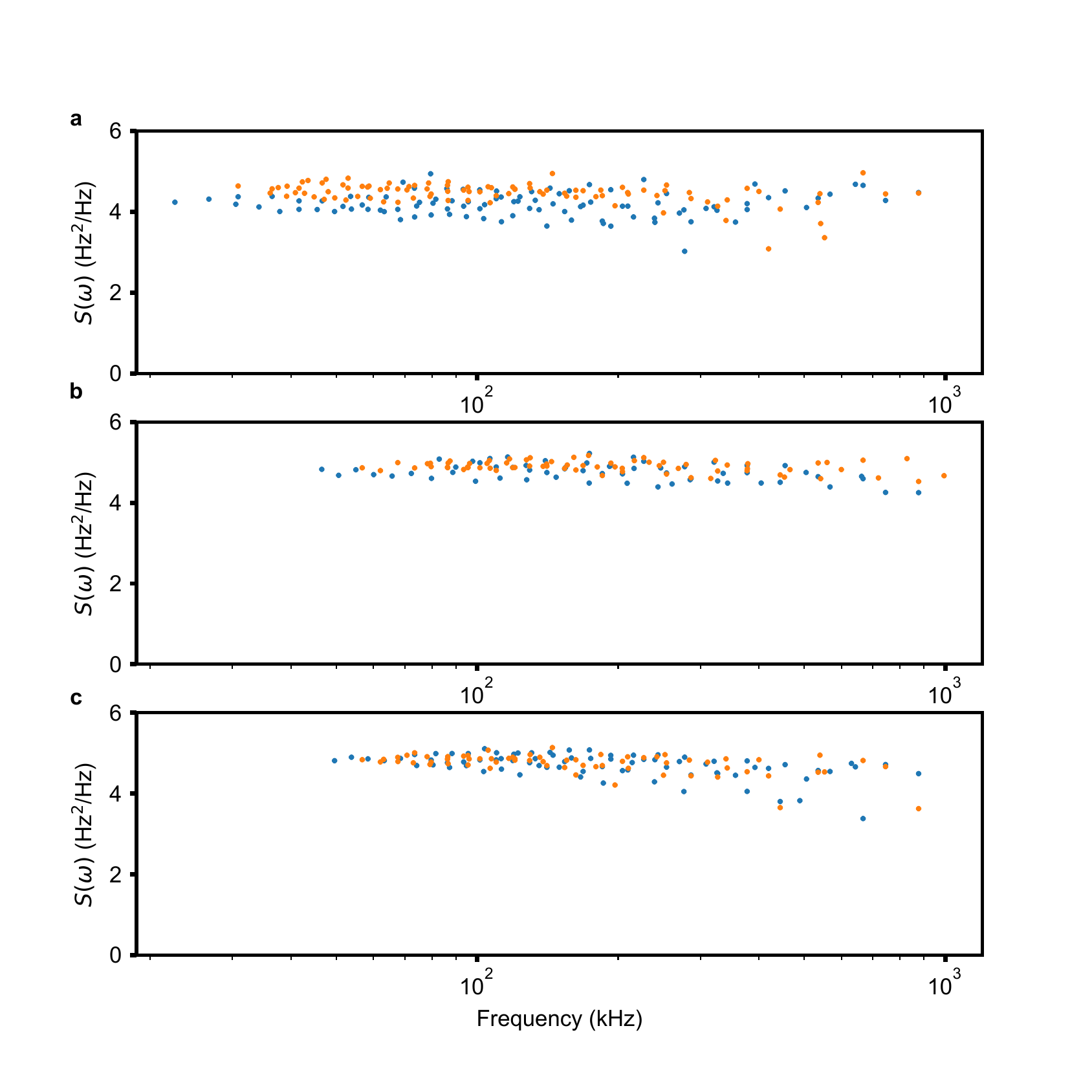}
    \caption{Spectral decomposition for sample of NV centers in \bosample{} before (blue) and after (orange) 650$^\circ$C vacuum anneal. Depths of NV centers are \textbf{a} 10.25 nm, \textbf{b} 9.72 nm, \textbf{c} 10.42 nm}
    \label{fig:spectraldecomp}
\end{figure}

We apply this procedure to the decoherence curves of select NV centers in \bosample{} before and after a 650$^\circ$C vacuum anneal to investigate the change in the noise spectrum due to annealing, and find that it is not significantly affected by the annealing procedure. These results are presented in Fig.~\ref{fig:spectraldecomp}. NV centers in this sample displayed reduced DEER signals after annealing, and so the absence of change in the noise spectrum indicates that the surface spins do not contribute significantly to the surface noise bath in this sample at the frequencies probed.

\subsection{Atomistic model for surface spins on (100) diamond surface}

X-ray spectroscopy data indicate that carbon sp$^3$ dangling bonds are a likely source of surface spins (Section \ref{sec:xrayspec}). This is a surprising result as surface sp$^3$ dangling bonds (DBs) appear to be chemically stable in ambient conditions, but are expected to be highly reactive in atmosphere. In order to set up a plausible model, we assume that some structural disorder is left on the surface after oxygenation, so the surface is not atomically smooth.

We propose a model to explain the existence and chemical stability of surface spins on the (100) diamond surface as sp$^3$ DBs at a (111) step edge. A (111) surface is essential to create a single dangling bond as, in contrast to a (100) surface, each carbon atom under a (111) surface possesses three bonds downward and one bond upward, from which it is possible to form a single sp$^3$ dangling bond defect with minimal change of the diamond lattice. In this model, all C(*) atoms are saturated with OH and the total spin of these systems is S=1/2. We also propose that the (111) facet, which is naturally generated at a step edge of (100) crystalline diamond surfaces, can sterically protect these spinful defects.

\begin{figure}[H]
    \centering
    \includegraphics[width=5in]{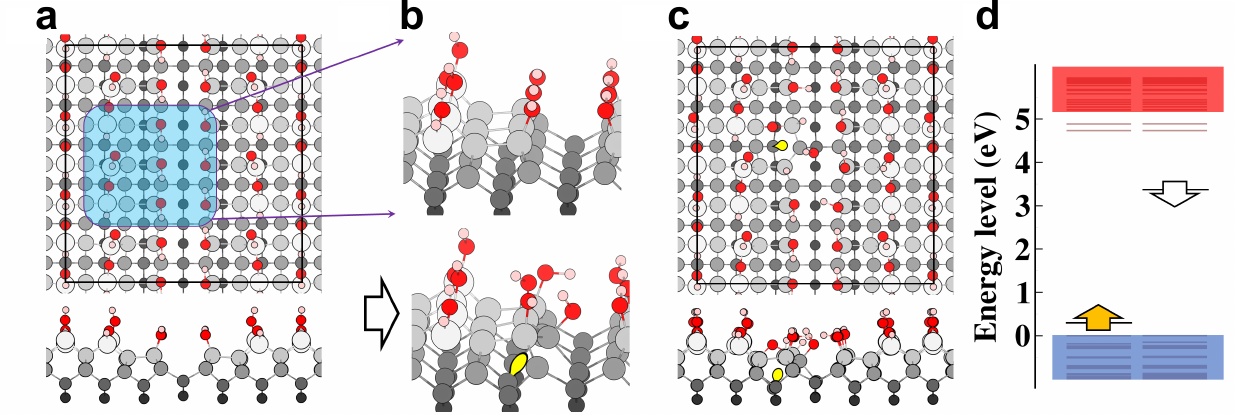}
    \caption{\textbf{Surface spin model.} \textbf{a}, The trench structure of (100)-6$\times$6 diamond surface. Surface termination is O/H/OH type, as in \cite{kaviani2014proper}, and step carbon atoms are saturated by OH. The red and pink balls are hydrogen and oxygen atoms, respectively. The grey balls are carbon atoms, the darker color and smaller size indicate deeper position away from the surface. For the sake of clarity, only the first five layers are presented in the structure plots. \textbf{b}, Illustration of the formation of a surface spin at the step edge. The yellow lobe represents the surface spin. \textbf{c}, A plausible structure model of a surface spin on the trench of the diamond (100) surface. \textbf{d}, The energy level plot of the modeled surface spin. The valence and conduction bands are depicted as blue and red regions, respectively. The orange and white arrows represent spin up and spin down channels, respectively. The spin-polarization energy is high for the dangling bond. We note that the single particle Kohn-Sham levels do not directly refer to the ionization energies of the dangling bond. The surface states are located at ~0.3 eV below the conduction band minimum. The valence band maximum positions are aligned to zero.}
    \label{fig:surfacespinmodel}
\end{figure}

It has been previously reported that O/H/OH mixed termination of (100) diamond minimally introduces defect levels to the bandgap \cite{kaviani2014proper}, so we use this surface as our starting model. In order to mimic the local disorder at the surface, we create two step edges forming a trench configuration \cite{chadi1987stabilities}, as shown in Fig.~\ref{fig:surfacespinmodel}(a). A previous study \cite{yang2008bond} indicated that such a trench structure can be energetically more favorable than the flat structure on the diamond (100) surface. We carried out first principles calculations on this surface model using the plane-wave based Vienna \textit{Ab Initio} Simulation Package \cite{kresse1996efficient}. The projector augmented-wave \cite{blochl1994projector} method is used to represent the electron-ionic core interactions. For the exchange-correlation functionals, we used spin-polarized gradient-corrected Perdew-Burke-Ernzerhof (PBE) functional \cite{perdew1996generalized} for structure optimization and Heyd-Scuseria-Ernzerhof (HSE06) \cite{heyd2003hybrid} hybrid functional for electronic structure calculations that would be capable of providing correct defect levels and defect-related electronic transition within ~0.1 eV to experiments \cite{gali2009theory}. A cutoff energy of 370 eV resulted in an equilibrium lattice parameter of diamond of 3.570 Å, which agrees with the experimental value of 3.567 \AA. We use a 6 $\times$ 6 supercell to simulate the trench models. The thickness of the vacuum layer is more than 10 \AA. There are eleven carbon layers, the surface seven layers are allowed to be fully relaxed until the forces are below 0.01~eV/\AA{} and the bottom four layers are fixed at their bulk positions. The k-point sampling is $\Gamma$ only, which is sufficient to map the Brillouin-zone.

In the starting model, a single carbon sp$^3$ dangling bond is introduced as the hypothetical source of the S=1/2 spin defect. The carbon atoms at the atomic step are more reactive because they do not possess ideal bonding configuration, and thus may be attacked by a radical such as –OH groups present in the oxygenation process. We simulated this situation as shown in Fig.~\ref{fig:surfacespinmodel}(b). First principles calculations showed that the surface carbon atom attacked by the –OH group spontaneously breaks a bond with a neighbor carbon atom beneath the surface, creating an sp$^3$ DB with S=1/2 spin as respectively shown in Fig.~\ref{fig:surfacespinmodel}(c) and (d). This near-surface carbon sp$^3$ DB resides in the fourth atomic layer referred to the topmost carbon atom layer which is topographically protected by the direct interaction with species from the gas phase. As expected, the sp$^3$ DBs are electrically and optically active; the latter may be achieved by photo-ionization to the negative charge state. The actual photo-ionization energy threshold may depend on the interaction with other defects at the surface or beneath the surface.

\section{Additional surface spin measurements}
\label{sec:surfacespinmeasurements}

\subsection{Coherently coupled surface spins}

\begin{figure}[H]
    \centering
    \includegraphics[width=5in]{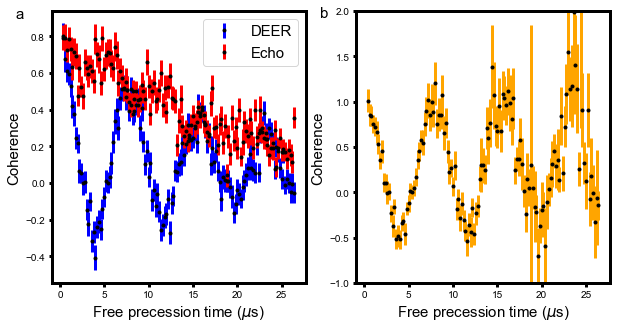}
    \caption{\textbf{Coherently coupled surface spin.} \textbf{a}, Raw DEER (blue) and echo (red) data from an NV center that is coherently coupled to a surface spin. \textbf{b}, The raw DEER signal divided by the echo signal.}
    \label{fig:CohCoup}
\end{figure}

Over all the samples investigated here, coherently coupled surface spins are rarely observed. One example of this rare occurrence is shown in Fig.~\ref{fig:CohCoup}. Based on the coupling strength, it is 6-7 nm from the NV center, which is consistent with the 7 nm depth as measured by proton NMR. However, we also note that it cannot be conclusively shown with our methods that this defect exists at the surface. This particular spin survived multiple acid cleanings and surface treatments and so may be deeper than the bath spins probed in the remainder of this work. There may exist several types of defects with an observable DEER signal, some of which are found below the diamond surface \cite{Grinolds2014}.

\subsection{Surface spin linewidth}
\label{linewidth}
\begin{figure}[H]
    \centering
    \includegraphics[width=5in]{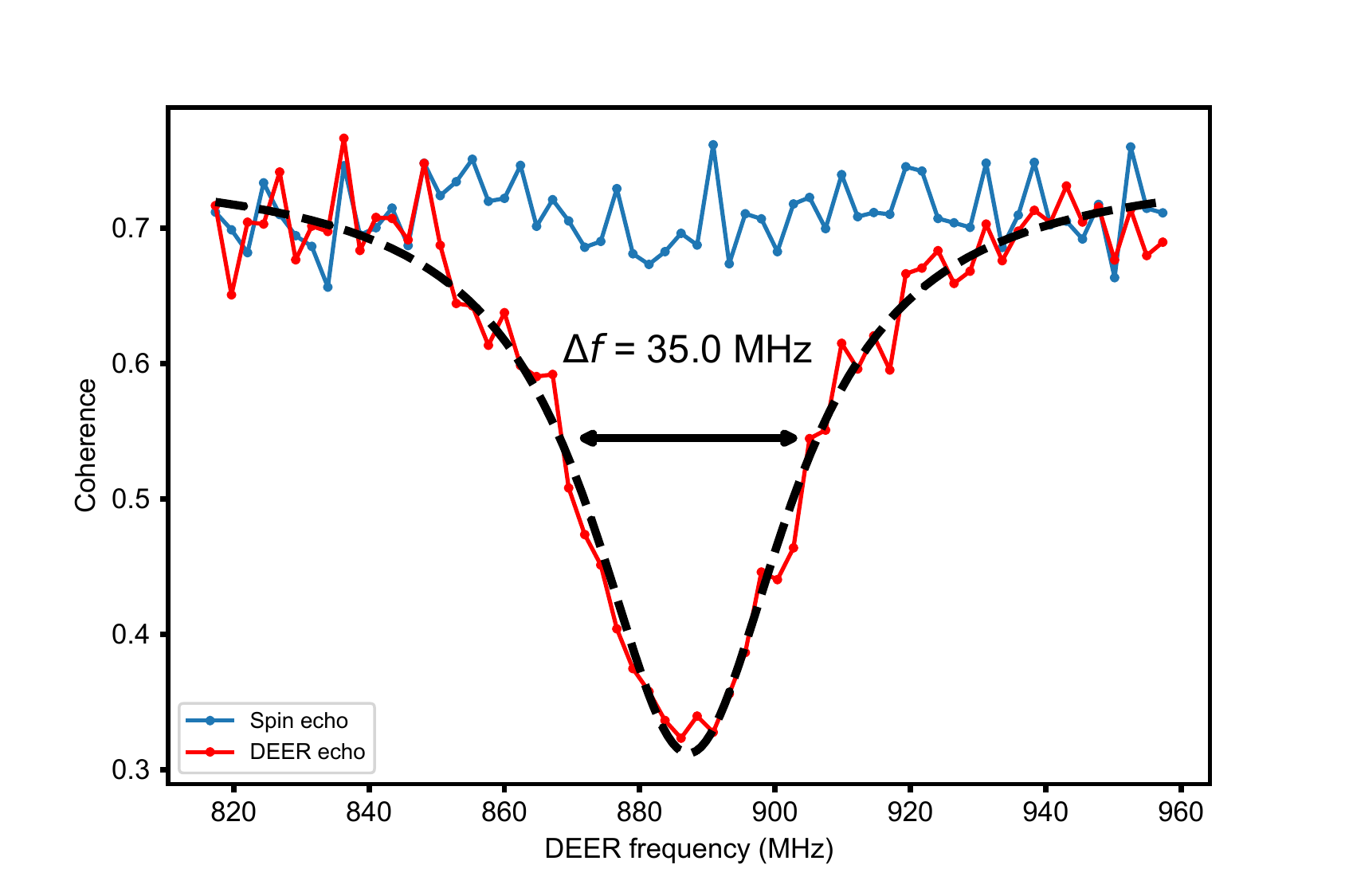}
    \caption{\textbf{DEER frequency sweep}
    }
    \label{fig:FrequencySweep}
\end{figure}

Another important parameter of the surface spin bath is the linewidth of the DEER transition, as this provides information about the disorder on the surface experienced by the surface spins. We measure the linewidth by performing a DEER pulse sequence while sweeping the frequency of the surface spin $\pi$ pulse. A typical dataset, with a Lorentzian fit, is shown in Fig.~\ref{fig:FrequencySweep}. We note that this value of 35 MHz is slightly larger than the $\approx 20 $ MHz linewidths reported in the literature \cite{bluvstein2019extending, Mamin2012, Grinolds2014, Sushkov2014}, which may be a result of power broadening. For the following discussions this discrepancy will not be relevant, however.

\subsection{Surface spin \texorpdfstring{T$_{2,\mathrm{Rabi}}$}{T2Rabi}}

\begin{figure}[H]
    \centering
    \includegraphics[width=5in]{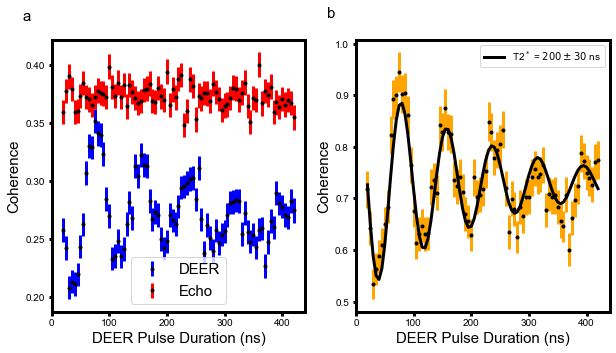}
    \caption{\textbf{DEER Rabi measurement.}
    \textbf{a} We perform a microwave pulse of varying length on the surface spins during the DEER sequence and \textbf{b} fit the normalized signal to a decaying exponential envelope to extract a coherence decay time.}
    \label{fig:DEERRabi}
\end{figure}

As an additional measurement to bound the disorder of the surface spins, we sweep the duration of the microwave pulse performed on the surface spins during the DEER sequence. We fit the resulting curve to a cosine with an exponentially decaying envelope, and find a characteristic decay time of 200 $\pm$ 30 ns (see Fig.~\ref{fig:DEERRabi}).

\subsection{Measured surface spin \texorpdfstring{$T_1$}{T1} relaxation times}
\label{T1section}
\begin{figure}[H]
    \centering
    \includegraphics[width=5in]{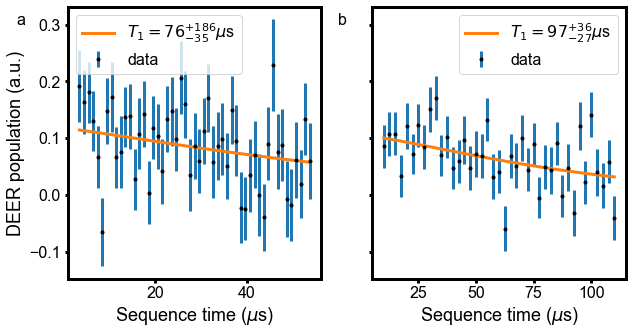}
    \caption{\textbf{Measured T1 of surface spins using correlation sequence.} Relaxation time $T_1$ for surface spins measured by two different NV centers \textbf{a} and \textbf{b}, using the correlation sequence described in \cite{Sushkov2014}.
    }
    \label{fig:ssT1}
\end{figure}

\begin{figure}[H]
    \centering
    \includegraphics[width=5in]{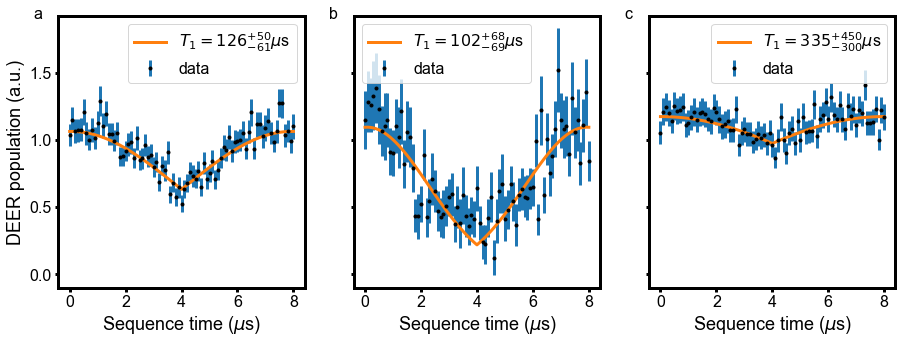}
    \caption{\textbf{Measured T1 of surface spins by bath correlation measurement.} Relaxation time $T_1$ for surface spins measured by three different NV centers, depicted in \textbf{a}, \textbf{b}, and \textbf{c}, using the sequence described in \cite{Mamin2012}. \textbf{a} and \textbf{b} are from \bosample{} and  \textbf{c} is from \NF{}.
    }
    \label{fig:RugarT1}
\end{figure}

The surface spin $T_1$ relaxation times were measured on several NV centers in \elanasample{} using a correlation spin sequence \cite{Sushkov2014, laraoui2013high}. Results can be seen in Fig.~\ref{fig:ssT1}. Additionally, for \bosample{}, a sampling of surface spin $T_1$ times were measured by sweeping the time of the DEER microwave pulse within the echo sequence \cite{Mamin2012}. Extracted $T_1$ times were consistent with those measured in correlation spectroscopy (see Fig.~\ref{fig:RugarT1}). These $T_1$ times were significantly longer than the length of the DEER sequences performed in the main text.

\section{Derivation of DEER signal for static and hopping surface spins}
\label{sec:derivation}
\subsection{System description}
We consider a single NV center coupled to a 2D spin bath. We assume the 2D spin bath to be located at $z=0$, in the $\hat{x}-\hat{y}$ plane, and the NV center to be located at $z=-d_{\rm{NV}}$. Consistent with a global magnetic field applied along the NV center quantization axis and for [100] diamond, we assume that all spins are oriented in a direction $54.7^\circ$ from the $\hat{z}$-axis. The Hamiltonian of the system with NV center spin $S$ with resonant frequency $\omega_s$, and surface spins $I_i$ with Larmor frequencies $\omega_i$ can be written as
\begin{align}
\label{eq:system_hamiltonian}
    \mathbf{H}=\omega_S \mathbf{S}^z + \sum_i \omega_i \mathbf{I}_i^z + \sum_i J_i \mathbf{S}^z\mathbf{I}_i^z
\end{align}
where we have assumed that the interaction between the NV center and surface spins takes an Ising form, since they have different resonance frequencies. The interaction strengths $J_i$ are assumed to be dipolar interactions, and we will explicitly calculate their distributions below.

In the above equation, we have also neglected the interactions between the surface spins, which may be justified for two reasons: first, local disorder fields experienced by the surface spins that cause their resonance frequencies to vary (see Section \ref{linewidth}, as well as \cite{Grinolds2014}), and this will suppress spin-exchange processes; second, the Ising interaction component will not change the $I^z$ component of each spin, and thus it will not affect the NV center spin signal. Note that we have also assumed a quasi-static bath for now and neglect the finite $T_1$ relaxation time of the spins. Spin-exchange processes between bath spins could be approximately treated in a similar fashion as a finite $T_1$ time. For completeness, we present some results from simulations that include interactions in section \ref{interactions}.

We now calculate the DEER signal expected for an NV center in the case of static surface spins and of surface spins undergoing configurational averaging.

\subsection{Derivation of DEER coupling for dense static spins}
The combined NV center-surface spin Hamiltonian is given by Eq.~\eqref{eq:system_hamiltonian}. In the rotating frame \cite{Slichter1990}, this simplifies to
\begin{equation}
    \mathbf{H} = \gamma_e \mathbf{S}^zB_z^{\rm{DEER}}
\end{equation}

\noindent where $\gamma_e$ is the electron gyromagnetic ratio and $B_z^{\rm{DEER}}$ is the total $z$ component of the magnetic field due to the surface spins at the location of the NV center. We now calculate the signal measured by an NV center during a DEER spin echo pulse sequence, in the case that other decoherence mechanisms affecting the NV center are negligible.

After spin initialization, the NV center spin starts out in the spin state $\ket{0}$ (m$_{\rm{s}}$ = 0), and the surface spins have an initial state vector $\ket{I} = \ket{I_1, I_2...I_n}$, where $\mathbf{I}^z_i\ket{I_i} = \pm \hbar/2\ket{I_i}$. The initial spin states are random for each repetition of the experiment. Then we apply a $\pi/2$ pulse to the NV center to put the system in the superposition state

\begin{equation}
\ket{\psi_{NV}(t = 0)} = \frac{1}{\sqrt{2}}\left(\ket{0} + \ket{1}\right)\otimes\ket{I}
\end{equation}

The NV center spin then evolves for a time $\tau/2$ under the influence of the magnetic field produced by the surface spins. This field is given by the summation of all dipole fields from the $n$ spins:

\begin{equation}
B_z^{\rm{DEER}}(t) = \frac{\mu_0}{4\pi}\sum_i \frac{3u_z^i(\vec{m^i}\cdot\vec{u^i}) - m_z^i}{\left|r^i\right|^3}
\end{equation}

\noindent where $\vec{u}$ is a unit vector from the NV center to spin $i$, $r^i$ is the distance to the $i$th spin, and $m^i = -\frac{\gamma_e\hbar}{2} \vec{\sigma^i}$, and $\sigma^i$ is the Pauli matrix for spin $i$.

After a time $\tau/2$, the NV center spin will be in a state

\begin{equation}
\ket{\psi_{NV}(t = \frac{\tau}{2})} = \frac{1}{\sqrt{2}}\left(\ket{0}\otimes\ket{I}\right) + \frac{1}{\sqrt{2}}\left(\ket{1}\otimes e^{-i\sum_i \phi_1^i}\ket{I}\right)
\end{equation}

\noindent where

\begin{equation}
\phi^i_1 = -\gamma_e^2 \hbar \frac{\mu_0}{8\pi\hbar}\frac{1}{\left|r^i\right|^3}(3u^i_zu^i_z - 1)I_i\frac{\tau}{2}
\end{equation}
\noindent and $I_i$ is either +1 or -1 for each spin, but it is initialized randomly for each experiment.

Next we apply a resonant $\pi$ pulse to the NV center and surface spins simultaneously, which has the effect of swapping $\ket{0}\leftrightarrow \ket{1}$ and also swapping $\ket{I_i}\rightarrow\ket{-I_i}$. The state is now

\begin{equation}
\ket{\psi_{NV}(t = \frac{\tau}{2})} = \frac{1}{\sqrt{2}}\left(\ket{1}\otimes \ket{-I}\right) + \frac{1}{\sqrt{2}}\left(\ket{0}\otimes e^{-i\sum_i \phi_1^i}\ket{-I}\right)
\end{equation}

After another free evolution period of $\tau/2$, we have

\begin{equation}
\ket{\psi_{NV}(t = \tau)} = \frac{1}{\sqrt{2}}\left(\ket{1}\otimes e^{-i\sum_i\phi_2^i}\ket{-I}\right) + \frac{1}{\sqrt{2}}\left(\ket{0}\otimes  e^{-i\sum_i\phi_1^i}\ket{-I}\right)
\end{equation}

\noindent where

\begin{equation}
\phi^i_2 = +\gamma_e^2 \hbar \frac{\mu_0}{8\pi\hbar}\frac{1}{\left|r^i\right|^3}(3u^i_zu^i_z - 1)I_i\frac{\tau}{2} = -\phi^i_1
\end{equation}

The final $\pi/2$ pulse on the NV center leads to

\begin{equation}
\ket{\psi_{NV}(t =\tau)} = \frac{1}{2}\left(\ket{1} - \ket{0}\right)\otimes e^{-i\sum_i\phi_2^i}\ket{-I} + \frac{1}{2}\left(\ket{0} + \ket{1}\right)\otimes e^{-i\sum_i\phi_1^i}\ket{-I}
\end{equation}

Measuring in the $\ket{0}$ basis then gives
\begin{align}
\left|\braket{0\mid \psi_{NV}}\right|^2 =& \left|-1/2e^{-i\sum_i\phi_2^i} + 1/2e^{-i\sum_i\phi_1^i}\right|^2\\
=& 1/2\left(1 + \cos\sum_i\Delta\phi^i\right)
\end{align}

\noindent where
\begin{equation}
\Delta\phi^i  = \gamma_e^2\hbar\frac{\mu_0}{8\pi\hbar}\frac{1}{\left|r^i\right|^3}(3u^i_zu^i_z - 1)I_i\tau
\end{equation}

If there is only one nearby dark spin coupled to the NV center, we will see full contrast oscillations at the coupling frequency of the surface spin.

The more typical situation is one in which the NV center senses many dark spins. At each experimental repetition, the NV center senses a different total magnetic field depending on the random spin states of the surface spins. The total signal is the average of this Bernoulli random process for many trials. For a large number of trials, this can be approximated by a Gaussian process, e.g. $\braket{\cos\phi} = e^{-\braket{\phi^2}/2}$. This will only be true if the NV center is coupled to sufficiently many spins with comparable coupling strengths, and in the limit of a dilute spin bath the Gaussian approximation will fail because of strong trial-to-trial fluctuations in the strength of dipolar coupling between the NV center and the dark spins, which lead to a distinctly non-Gaussian distribution of acquired phases \cite{klauder1962spectral, dobrovitski2008decoherence}.

Computing the average of squares we find

\begin{equation}
\braket{\Delta\phi^2} = \left(\frac{\tau \gamma_e^2\mu_0\hbar}{8\pi}\right)^2\sum_i\frac{1}{\left|r^i\right|^6}(3u^i_zu^i_z - 1)^2
\end{equation}

If the surface spins are dense enough that the average spin-spin spacing is much smaller than the NV center depth, $d_{\rm{NV}}$, the sum can be approximated as an integral over a continuous density of surface spins, $\sigma$. We can write $r = \sqrt{a^2 + d_{\rm{NV}}^2}$ and $u_z = \frac{1}{r}(d_{\rm{NV}}\cos\theta + a\sin\theta)\cos\alpha$ and calculate

\begin{align}
\sum_i\frac{1}{\left|r^i\right|^6}(3u^i_zu^i_z - 1)^2\approx& \int_0^{2\pi}\int_0^{\infty} \frac{da d\alpha a}{(d_{\rm{NV}}^2 + a^2)^{3}}(\frac{3}{a^2 + d_{\rm{NV}}^2}(d_{\rm{NV}}\cos\theta + a\sin\theta)^2\cos^2\alpha - 1)^2\\
=& \frac{3\pi}{8d_{\rm{NV}}^4}
\end{align}

\noindent for $\theta = 54.7^\circ$, the angle of NV centers in (100) diamond with respect to the surface. Converting the population to the coherence, we finally obtain

\begin{equation}\label{eq:Gauss}
\left<S(\tau)\right> =\exp\left(-\left(\frac{\mu_0}{4\pi}\right)^2\frac{3\pi\gamma_e^4\hbar^2 \sigma}{64 d_{\rm{NV}}^4}\tau^2\right)
\end{equation}

\subsection{Calculation of NV coherence decay for configurationally averaged surface spins}

Based on experimental observations of a small stretched exponent and the absence of strongly coupled spins, we calculate the case where surface spins can change positions between measurements, effectively allowing a configurational averaging when evaluating the coherence decay.

Since configurational averaging can result from random spin hopping, we can make use of the methods in Ref.~\cite{fel1996configurational,Abragam1961} to calculate the coherence decay. The intuition of the calculation is that, since all interactions take the Ising form, we can simply add up the phases. It turns out that the resulting expression, after being written in an exponential form, can be factorized into a product of cosines corresponding to the individual phase accumulations, which can then be integrated over different spatial configurations. We will not reproduce the full details here, but simply note that the same argument should also apply to a spin located at some distance external to the spin bath, and one simply needs to replace the interaction strength by the corresponding one for the external spin.

Thus, we need to calculate the interaction strength between the NV center located at $(0,0,-d_{\rm{NV}})$ and a spin located at position $(r\cos\alpha,r\sin\alpha,0)$. The quantization axis is assumed to be pointing in $\hat{n}=(\sin\theta,0,\cos\theta)$, where $\theta=54.7^\circ$, such that $\cos^2\theta=1/3$. Taking the inner product, we have that the interaction strength is
\begin{align}
\frac{\mu_0}{4\pi}\frac{\gamma_e^2\hbar t}{2(r^2+d_{\rm{NV}}^2)^{3/2}}\left(\frac{3(r\cos\alpha\sin\theta+d_{\rm{NV}}\cos\theta)^2}{r^2+d_{\rm{NV}}^2}-1\right)=\frac{\mu_0}{4\pi}\frac{\gamma_e^2\hbar t(2r^2\cos^2\alpha-r^2-2\sqrt{2}rd_{\rm{NV}}\cos\alpha)}{2(r^2+d_{\rm{NV}}^2)^{5/2}}
\end{align}

Transcribing the expressions found in Ref.~\cite{fel1996configurational} with these interactions, we find that the coherence decay should satisfy
\begin{align}
\langle S(t)\rangle_c &=\exp\left\{\sigma\int_0^{2\pi}d\alpha\int_0^\infty rdr\times [\cos \frac{\mu_0\gamma_e^2\hbar \tau(2r^2\cos^2\alpha-r^2-2\sqrt{2}rd_{\rm{NV}}\cos\alpha)}{8\pi(r^2+d_{\rm{NV}}^2)^{5/2}}-1]\right\}\nonumber\\
&=\exp\left\{\sigma d^2\int_0^{2\pi}d\alpha\int_0^\infty xdx\times [\cos \frac{\mu_0\gamma_e^2\hbar \tau}{8\pi d_{\rm{NV}}^3}\frac{(2x^2\cos^2\alpha-x^2-2\sqrt{2}x\cos\alpha)}{(x^2+1)^{5/2}}-1]\right\},
\label{eq:integral}
\end{align}
where we have written $x=r/d_{\rm{NV}}$.

We would like to note that here $\mu_0\gamma_e^2\hbar \tau/8\pi d_{\rm{NV}}^3$ always appears together in the integral, which means that the shape of the curve  should be the same regardless of the depth $d_{\rm{NV}}$. Changing $d$ only leads to a rescaling of the time at which changes in shape happen. However, note that since there is a $d_{\rm{NV}}^2$ in front, the timescale involved in the stretched exponential decay can change.

While the above expression does not appear to be easily integrated, we can take appropriate limits to understand its behavior. First, let us consider the limit where $\tau$ is small, such that $\mu_0\gamma_e^2\hbar \tau\ll 8\pi d_{\rm{NV}}^3$. In this case, intuitively, the spins far away will not have the time to interact yet, so we are dominated by contributions from small $r$. In this case, the argument of the cosine in Eq.~(\ref{eq:integral}) is small, such that we can perform a Taylor expansion to obtain
\begin{align}
\langle S(t)\rangle_c &=\exp\left\{\sigma d_{\rm{NV}}^2\int_0^{2\pi}d\alpha\int_0^\infty xdx\times [-\frac{1}{2}\left(\frac{\mu_0\gamma_e^2\hbar \tau}{8\pi d_{\rm{NV}}^3}\frac{( 2x^2\cos^2\alpha-x^2-2\sqrt{2}x\cos\alpha)}{(x^2+1)^{5/2}}\right)^2]\right\}\nonumber\\
&=\exp\left[-\frac{3\pi\sigma d_{\rm{NV}}^2}{16}\left(\frac{\mu_0\gamma_e^2\hbar \tau}{8\pi d_{\rm{NV}}^3}\right)^2\right].
\label{eq:shorttime}
\end{align}

Thus, at short times, we recover the Gaussian decay in Eq.~\ref{eq:Gauss}.

Next, let us consider the limit where $\tau$ is large, such that $\mu_0\gamma_e^2\hbar \tau\gg 8\pi d_{\rm{NV}}^3$. Intuitively, the time is long enough such that the external spin sees the influence of all spins far away, and the situation will be similar to a 2D spin bath. In this case, we will have contributions from a large range of values of $x$. Due to the $x$ factor outside the parenthesis however, coming from the volume element in the Jacobian, the main contribution to the integral will be coming from values of $x$ that are large, and thus we may approximate the fraction as
\begin{align}
\frac{(2x^2\cos^2\alpha-x^2-2\sqrt{2}x\cos\alpha)}{(x^2+1)^{5/2}}\approx \frac{(2x^2\cos^2\alpha-x^2)}{x^5}=\frac{2\cos^2\alpha-1}{x^3}.
\end{align}

Explicitly performing the integral in this limit gives

\begin{align}
\langle S(t)\rangle_c=\exp\left(-\frac{9\sqrt{\pi}\Gamma(\frac{11}{6})\sigma}{5}\left(\frac{\mu_0\gamma_e^2\hbar \tau}{8\pi}\right)^{2/3}\right),
\label{eq:longtime}
\end{align}
which reproduces the $\tau^{2/3}$ scaling in this limit. Note also that in this limit, $d_{\rm{NV}}$ drops out of the problem. The intuition is again that the spin sees everything out to infinity, and the only spatial scale left in the problem is an average distance between the surface spins.

\subsection{Mean nearest neighbor separation}
In the main text, an average spin-spin separation, $\ell_{SS}$, is calculated from a fitted density of surface spins, $\sigma$. From dimensional analysis, these quantities are related by
\begin{equation}
\ell_{SS} \propto \frac{1}{\sqrt{\sigma}}
\end{equation},
\noindent however the constant of proportionality is ambiguous. In the main text, we specifically are referring to the mean value of the distribution of nearest neighbor distances for a randomly distributed layer of spins with density $\sigma$. In this case the constant of proportionality is $1/2$, and for completeness we derive this result below.

We can obtain this by considering the probability of finding a neighboring spin a distance $r$ away. The distribution function, $w(r)dr$ = P(no neighbor < r)$\times$P(neighbor between $r$ and $r + dr$):
\begin{equation}
w(r)dr = \left(1 - \int_0^rw(r')dr'\right)\times(2\pi r dr\sigma)
\end{equation}
\noindent{}Taking the derivative gives
\begin{equation}
\frac{dw(r)}{dr} = 2\pi \sigma \left(1 - \int_0^rw(r')dr'\right) - w(r)\times(2\pi r\sigma)
\end{equation}
\noindent{}Substituting into the first equation we can eliminate the integral
\begin{equation}
\frac{dw(r)}{dr} = \frac{w(r)}{r} - w(r)\times(2\pi r\sigma)
\end{equation}
\noindent{}This differential equation has the solution
\begin{equation}
w(r) = Kr\exp\left\{-\pi r^2 \sigma\right\}
\end{equation}
Where $K$ is an integration constant, which from normalization must be $K = 2\pi\sigma$. So the nearest neighbor probability distribution function is
\begin{equation}\label{eq:nearestneighbordist}
w(r) = 2\pi\sigma r \exp\left\{-\pi r^2 \sigma\right\}
\end{equation}

\noindent{}We then calculate the first moment of Eq.~\eqref{eq:nearestneighbordist}:
\begin{align}
\left<\ell_{SS}\right> =&\int_0^\infty rw(r)dr\\
=& 2\pi\sigma\frac{1}{4\pi\sigma^{3/2}}\\
=& \frac{1}{2\sqrt{\sigma}}\label{eq:nn}
\end{align}
\noindent{}Eq.~\eqref{eq:nn} is the expression used in the calculations of the Main Text.

\section{Effects of finite surface spin \texorpdfstring{$T_1$}{T1} time}
\label{sec:effectsoft1}
Here we calculate the effects of finite surface spin $T_1$ time on the DEER exponent $n$ in order to investigate whether the observed coherence decay behavior can be accounted for by surface spin relaxation.

In the presence of Gaussian-distributed noise amplitudes (as expected from a continuous density of surface spins), the quasistatic limit $T_1 \gg T_{2,\text{DEER}}$ gives a DEER decay exponent $n=2$, and the Markovian limit $T_1 \ll T_{2,\text{DEER}}$ gives a DEER decay exponent $n=1$. Quantitatively, the case of finite-$T_1$ time is calculated in \cite{Mamin2012} and is given by

\begin{equation}
    \left<S(\tau)\right> = \exp{\left[-\left(\Gamma_{\text{DEER}} \tau\right)^2 f(\tau,T_1)\right]},
    \label{eq:finiteT1}
\end{equation}

\noindent{}where

\begin{equation}
    f(\tau,T_1) = \frac{2 T^2_1}{\tau^2} \left[\frac{\tau}{T_1} - 1 + e^{-\tau/T_1}\right],
    \label{eq:foftau}
\end{equation}

\noindent{}is a dimensionless quantity that describes the effects of finite surface spin $T_1$ relaxation time. Note that in the limit of $\tau \ll T_1$ one finds $f(\tau,T_1) = 1$.

\begin{figure}[H]
\includegraphics[width=6.75in]{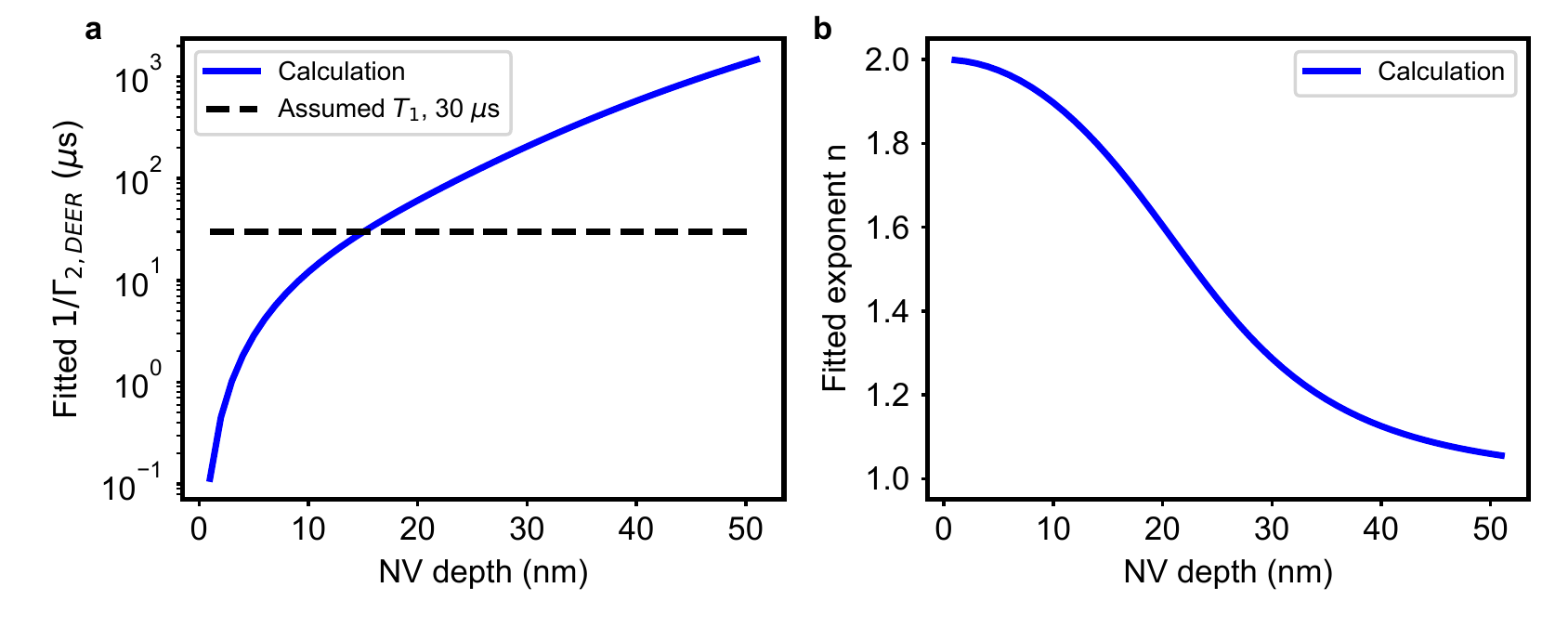}
\caption{\textbf{Depth scaling in the presence of finite surface spin $T_1$ time.} \textbf{a}, Calculated decay timescale $T_{2,\text{DEER}}$ with decay profile calculated using Eq.~\ref{eq:finiteT1} and fit to $\exp(-(\tau/T_{2,\text{DEER}})^n)$, assuming a finite surface spin $T_1$ time of $30~ \mu$s (black dashed line) and a density $0.005 / \text{nm}^2$. \textbf{b}, Fitted exponent $n$. Once $T_{2,\text{DEER}}$ becomes longer than the surface spin $T_1$ relaxation time, $n$ decreases from the infinite-$T_1$ limit of $n=2$ toward the Markovian limit of $n=1$. The calculated increase of $n$ for shallower NV centers is opposite of the observed trend vs depth in Fig.~2c of the Main Text, and suggests that the depth trend in Fig.~2c is not explained by the effects of finite relaxation time.}
\label{fig:ssfiniteT1}
\end{figure}

In Fig.~\ref{fig:ssfiniteT1} we fit Eq.~\ref{eq:finiteT1} to $\exp(-(\tau/T_{2,\text{DEER}})^n)$. As expected, for shallow NV centers $T_{2,\text{DEER}} \ll T_1$ and so we recover the quasistatic limit $n=2$, whereas for deeper NV centers $T_{2,\text{DEER}} \gg T_1$ and we recover the Markovian limit $n=1$. The trend of $n$ vs depth in Fig.~\ref{fig:ssfiniteT1}(b) is opposite from the trend observed in Fig.~2c of the Main Text, indicating that spin relaxation cannot account for our data.

\subsection{Exponents \texorpdfstring{$n$}{n} under the assumption of a noise spectral density}\label{n_derivation}

Here we show that the exponents $n$ from DEER exponential decays of the form $C(\tau) = \exp[-(\tau/T_2)^n]$ will have $n\geq1$ if the noise spectrum is flat ($n=1$) or monotonically decreasing as a function of frequency ($n>1$), assuming the noise spectral density formulation holds. The assumption of a monotonically decreasing noise spectrum is physically reasonable and consistent with experimentally observed noise spectra; accordingly, a likely explanation for the observed DEER exponents $n < 1$ is that the noise spectral density formulation does not hold, in particular because the noise amplitudes are not Gaussian-distributed, as is found in the case of discrete hopping spins.

We now show that $n \geq 1$ for a monotonically decreasing noise spectral density. We show this to be true for a general N-pulse qubit filter function $\frac{F_N(\omega T)}{\omega^2}$, defined to be continuous and non-negative across all frequencies, including Ramsey / DEER measurements with filter function $\frac{4 \sin^2(\omega T/2)}{\omega^2}$. Note that during a DEER sequence, if the surface spins are perfectly flipped during the NV center $\pi$ pulse, then the NV center filter function to the surface spins will be identical to that of Ramsey. The FID can then be described by Eq.~\eqref{eq:coherence}. The noise spectrum $S(\omega) \geq 0$, and for physically relevant noise sources is  monotonically decreasing, i.e. $\frac{dS(\omega)}{d\omega} \leq 0~ \forall \omega$, and with $S(\omega)$ and $\frac{dS(\omega)}{d\omega}$ defined over $\omega \in [0, \infty)$. We rewrite Eq.~\eqref{eq:spec} as

\begin{equation}
    \chi(T) = \frac{1}{\pi} T \left[\int_0^{\infty} S\left(\frac{x}{T}\right) \frac{F_N(x)}{x^2} dx \right],
    \label{chiwithg}
\end{equation}

\noindent where $x \equiv \omega T$ is a dimensionless integration variable. In order to probe the exponent $n$ associated with $\chi(T) = (T/T_2)^n$, we apply $(T \frac{d}{dT} - 1)$ to $\chi(T)$ in Eq.~\eqref{chiwithg}; using the product rule we find

\begin{equation}
    \left(T \frac{d}{dT} - 1\right) \chi(T) = \frac{1}{\pi} T^2 \left[\int_0^{\infty} \frac{dS(x/T)}{dT} \frac{F_N(x)}{x^2} dx \right].
    \label{dchi}
\end{equation}

Using  $\omega(T) = x / T$, we can evaluate

\begin{equation}
    \frac{dS(\omega(T))}{dT} = \frac{dS(\omega)}{d\omega} \frac{d\omega(T)}{dT} = \frac{dS(\omega)}{d\omega} \left(\frac{-x}{T^2}\right),
\end{equation}

Eq.~\ref{dchi} simplifies to

\begin{equation}
    T \frac{d\chi(T)}{dT} - \chi(T) = \frac{1}{\pi} \int_0^{\infty} \left(\frac{-dS(\omega)}{d\omega} \right) \biggr\rvert_{\omega = x/T} \frac{F_N(x)}{x} dx %
    \label{dchi2}
\end{equation}

We can rewrite the left-hand-side of Eq.~\ref{dchi2} by using $dT/T = d \ln T$, and since, by assumption, $\chi(T)$ is positive and differentiable at times $T>0$, we write

\begin{equation}
    T \frac{d\chi(T)}{dT} - \chi(T) = \chi(T) \left( \frac{1}{\chi(T)} \frac{d \chi(T)}{d \ln T} - 1 \right) = \chi(T) \left(\frac{d \ln \chi(T)}{d \ln T} - 1 \right)
    \label{dchilhs}
\end{equation}

So, Eq.~\ref{dchi2} can be written as

\begin{equation}
    \frac{d \ln \chi(T)}{d \ln T} = 1 + \frac{1}{\chi(T)} \frac{1}{\pi} \int_0^{\infty} \left(\frac{-dS(\omega)}{d\omega} \right) \biggr\rvert_{\omega = x/T} \frac{F_N(x)}{x} dx
    \label{dchi3}
\end{equation}

As defined, the right-hand-side is necessarily $\geq$ 1 for all times $T$ because $\chi(T) > 0 ~\forall T$,  $F_N(x) \geq 0 ~\forall x$ and $\frac{dS(\omega)}{d\omega} \leq 0 ~\forall \omega$. Accordingly,

\begin{equation}
    \frac{d \ln \chi(T)}{d \ln T} \geq 1
    \label{dchigtr}
\end{equation}

\noindent for all free precession times $T > 0$. And, if in some region of times $T$ the function $\chi(T)$ has a power-law form $(T/T_2)^n$, then

\begin{equation}
    n \geq 1
\end{equation}

Under these mathematical arguments, we cannot observe $n < 1$ for a monotonically decreasing noise spectral density, because the integral is shown to be $\geq 0$ for all times $T$. A monotonically increasing noise spectral density would give a negative integral in Eq.~\ref{dchi3} and therefore $n<1$, but this is physically unreasonable and inconsistent with observed noise spectra.

Further, note Eq.~\ref{dchigtr} holds for a general $\chi(T)$, and implies that the slope on a plot of $\ln{\chi(T)}$ versus $\ln(T)$ will be $\geq$ 1 for a monotonically decreasing noise spectral density. In the Main Text we plot $\ln{\chi(T)}$ versus $\ln(T)$ and find slopes less than 1, which implies that the noise spectral density formulation does not hold, in particular because the noise amplitudes are not Gaussian-distributed. The configurational averaging of surface spins results in a non-Gaussian distribution of noise amplitudes (i.e., for some shots there are many surface spins nearby and for some shots there are no surface spins nearby), which violates the requirements for the noise spectral density formulation and explains the exponents $n<1$ that we observe.

\subsection{Note about filter functions}
In the above derivation, it is assumed that the filter function used during a measurement is that of the Ramsey free induction decay (FID) due to the surface spins alone. Experimentally this is not what is measured due to non surface spin contributions to the FID. In the Main Text, we attempt to remedy this by dividing by a spin echo signal. However, depending on the form of the noise spectrum, this filter function may be different from what is desired.

To see this, we write the noise spectrum experienced by the NV center as a sum of two terms:

\begin{equation}
S(\omega) = S(\omega)_{\rm{SS}} + S(\omega)_{\rm{EE}}
\end{equation}

\noindent{}where $S(\omega)_{\rm{SS}}$ is the noise spectrum of the surface spins and $S(\omega)_{\rm{EE}}$ is the noise spectrum due to all other noise sources. The spin echo measurement probes the entire noise spectrum $S(\omega)$ with the spin echo filter function. In the spectral noise formalism the signal is given by

\begin{equation}
C(\tau) = \exp \frac{-1}{\pi} \int_0^\infty S(\omega) F^{SE}(\omega\tau) \frac{d\omega}{\omega^2}
\end{equation}

\noindent{}where $F^{SE}(\omega\tau)$ is the spin echo filter function, given by

\begin{equation}
F^{SE}(\omega\tau) = 8\sin^4 \frac{\omega\tau}{4}\,.
\end{equation}

The DEER measurement, in contrast, probes $S(\omega)_{\rm{SS}}$ with the spin echo filter function, but $S(\omega)_{\rm{EE}}$ is probed with the Ramsey filter function, $F^R(\omega)$, given by

\begin{equation}
F^{R}(\omega\tau) = 2\sin^2 \frac{\omega\tau}{2}\,.
\end{equation}

The DEER measurement signal is then given by

\begin{equation}
D(\tau) = \exp \frac{-1}{\pi} \int_0^\infty \left(S(\omega)_{\rm{SS}}F^{R}(\omega\tau) + S(\omega)_{\rm{EE}}SF^{SE}(\omega\tau)\right) \frac{d\omega}{\omega^2}\,.
\end{equation}

When we divide these two signals, what we actually measure is not the FID of the surface spins alone, but rather the quantity

\begin{equation}
S(\tau) = \exp \frac{-1}{\pi} \int_0^\infty S(\omega)_{\rm{SS}} \left(F^{R}(\omega\tau) - F^{SE}(\omega\tau)\right) \frac{d\omega}{\omega^2}\,.
\end{equation}

If the surface spins have a long relaxation time ($T_1^{DEER} > T_2^{NV}$), the second term is negligible due to the low weight of the filter function near $\omega = 0$. For fast relaxation times, the stretching factor of the quantity $S(\tau)$ does not go to 2/3, but rather to 0, and this occurs for $\tau > T_1^{DEER}$. This can be intuitively understood by the following argument: once the free precession interval exceeds $T_1^{DEER}$, there is no difference between the measurement with and without a $\pi$ pulse on the surface spins and so the two measurements become nominally identical. Our measurements of surface spin relaxation times (see Section \ref{T1section}) indicate that we are not in this regime, but for systems with much shorter relaxation times this additional transition should be observable.

\subsection{Effects of interactions}\label{interactions}
\begin{figure}[H]
\includegraphics[width=6.75in]{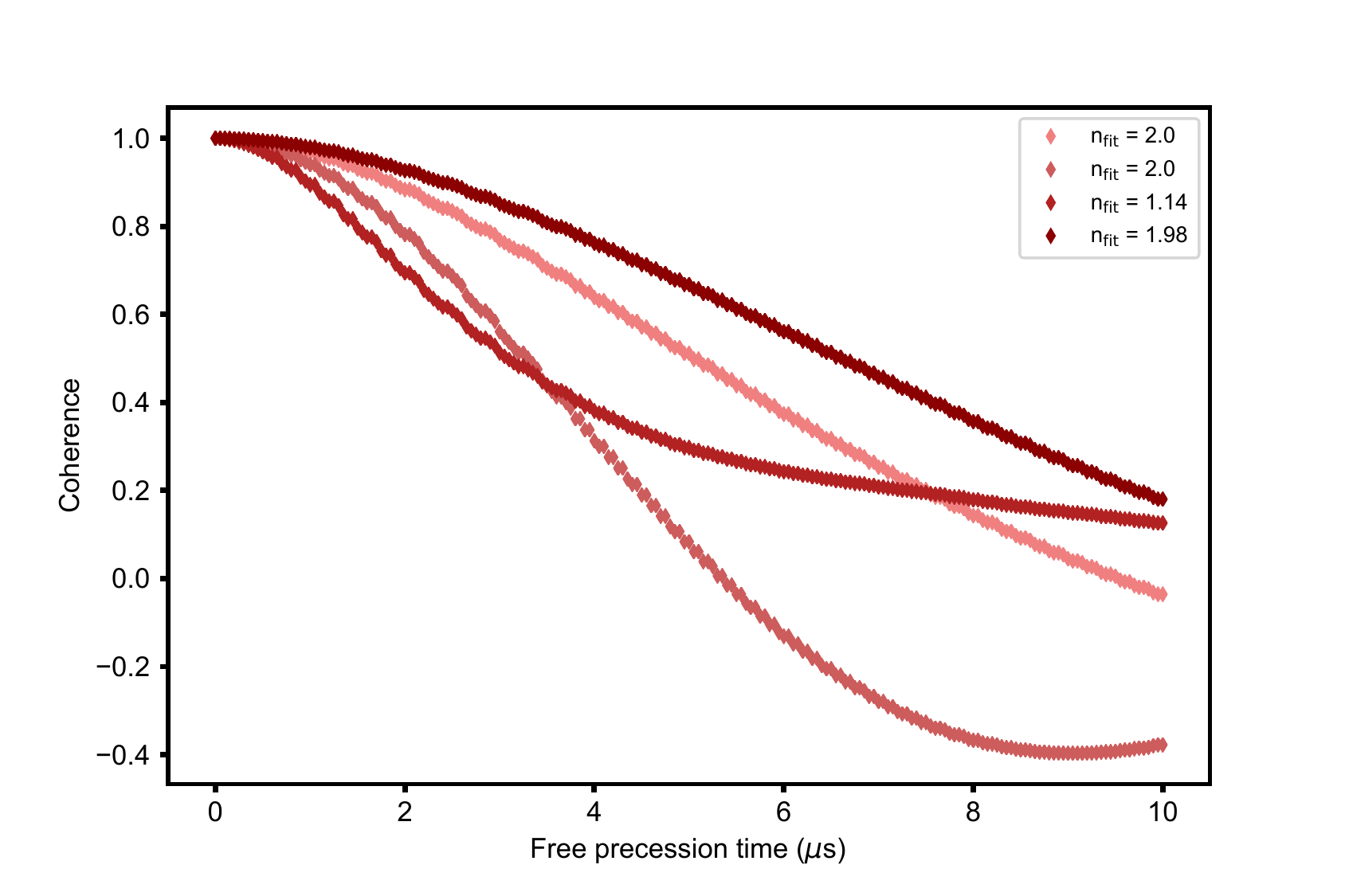}
\caption{\textbf{Simulations with incoherent interactions.} Example DEER curves for simulations allowing both flip-flops and spin lattice relaxation between surface spins. Each color represents a different randomly sampled surface spin configuration. Fitting these curves with a stretched exponential produces values of $1 < n < 2$. Some curves additionally show coherent oscillations if the coupling rate to the nearest spin exceeds the rate of incoherent decay.}
\label{fig:flipflopfig}
\end{figure}

Spin flips among surface spins can arise from environmental interactions and interactions between the surface spins (e.g. dipolar flip-flops). In order to determine whether flip-flops among static surface spins could reproduce the observed experimental results, we performed Monte Carlo simulations for $\sim 100$ random fixed spin configurations. Interactions between the spins were assumed to be incoherent due to the large mismatch between the measured disorder of surface spin systems ($1/(2\pi T_2^*)\approx 35$ MHz, see Section \ref{linewidth}) and the average nearest neighbor dipolar couplings at the inferred spin densities ($\approx 100$ kHz). In this regime, the probability of a spin exchange between spin $i$ and $j$ is given by \cite{Hall2016}

\begin{equation}
\label{eq:incoherentflipflop}
p_{i,j} = \frac{\omega^2_{i,j}}{\gamma}\delta t
\end{equation}
\noindent where $\omega_{i,j}$ is the dipolar coupling between spins $i$ and $j$ and $\gamma$ is the disorder. $\delta t$ is the time step of the simulation, chosen such that all $p_{i,j} < 1$. Additionally, spins are allowed to relax without spin exchange through spin-lattice relaxation at a rate $1/(2T_1)$. For a simulation, spins were given random coordinates, the all-to-all couplings were calculated, and the spins randomly initialized. At each time step, spins were allowed to flip or flip-flop based on draws from a Bernoulli distribution, and the phases for the entire time evolution were summed. Some results for simulation parameters $d_{\rm{NV}} = 5$ nm, $\sigma_{2D} = .005$ nm$^{-2}$, $T_1 = 50$ $\mu$s, and $\gamma = 100$ kHz are shown in Fig.~\ref{fig:flipflopfig}. For these parameters, $\sim20$\% of configurations lead to the observation of coherent oscillations in the DEER signal (whenever the nearest spin coupling is greater than the incoherent decay rate). No simulations with experimentally relevant parameters have resulted in stretching factor $n < 1$.

\section{Exponent transition for additional NV centers}
\label{sec:additionalNVs}
\begin{figure}[H]
    \centering
    \includegraphics[width=7in]{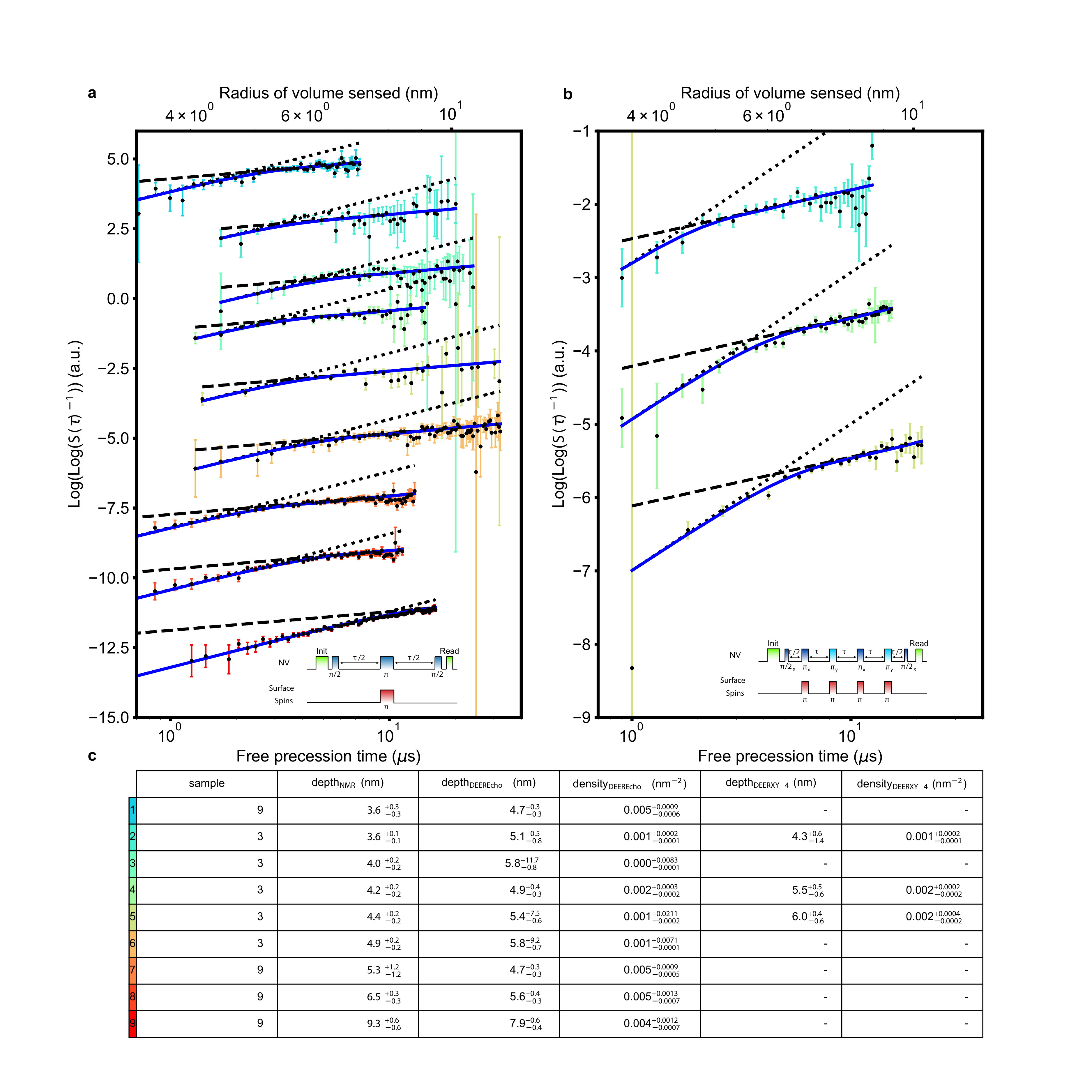}
    \caption{\textbf{Additional NV center measurements} \textbf{a}, $\log S(\tau)^{-1}$ for nine additional NV centers. Inset: pulse sequence used to measure NV centers in (a). Blue lines are fits to the hopping model in the Main Text, and dotted and dashed lines show slopes of 2 and 2/3, respectively. Y axis offset is arbitrary. \textbf{b}, DEER XY-4 data for four additional NV centers, plotted as in (a). Inset: DEER XY-4 pulse sequence. \textbf{c}, Surface spin densities and depths extracted from NV centers in (a) and (b). Depth$_{NMR}$ is the NV center depth extracted from fitting the proton NMR signal, while Depth$_{DEER}$ and Depth$_{DEERXY4}$ are the depths extracted from the hopping model in the Main Text. Error bars are one standard error.}
    \label{fig:additionaldata}
\end{figure}

In Fig.~\ref{fig:additionaldata}, we present additional finely sampled datasets from NV centers of various depths as in Fig.~3a. Again, we plot $\log S(\tau)^{-1}$ on a log-log scale so that the exponent, $n$, is given by the slope of the line. In addition to the DEER Echo type measurements presented in the Main Text, for some NV centers we performed a dynamical decoupling sequence (XY-4) on the NV center while applying multiple $\pi$ pulses to the surface spins, which in principle permits a longer interrogation of the surface spin FID before the NV center decoheres. The interpretation of this type of measurement is complicated by the short dephasing time of the surface spins and the longer total rotation time required for multiple $\pi$ pulses. The close agreement with $n=2/3$ at long times is striking nevertheless.

\section{Additional model analysis}
\label{sec:addtionalmodelanalysis}

\subsection{Analysis of two-exponent fits}

\begin{figure}[H]
    \centering
    \includegraphics[width=5in]{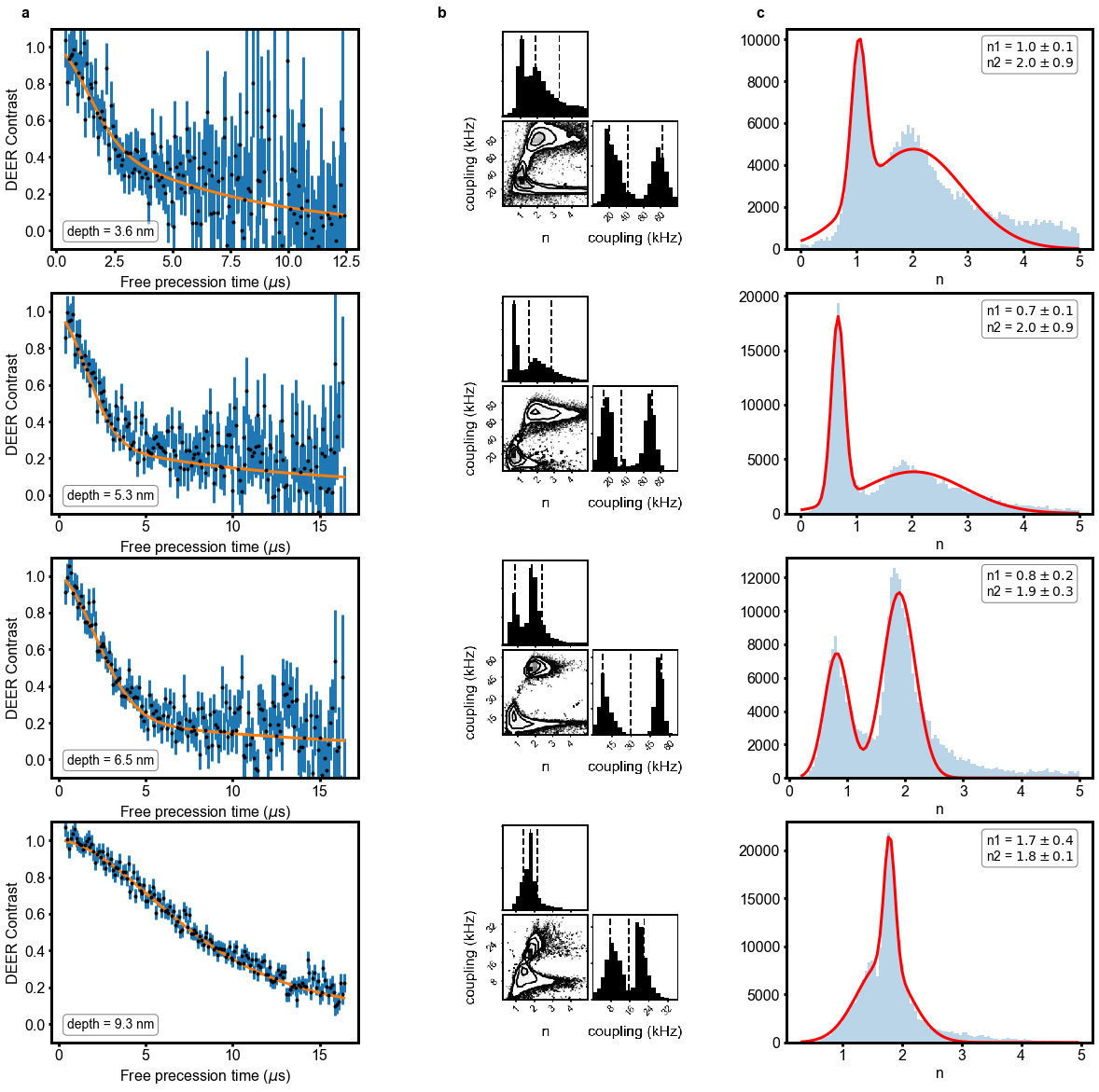}
    \caption{\textbf{Fits to a bi-exponential for four NV centers} \textbf{a}, Data for each of the four NV centers. The orange lines are best bi-exponential fits based on MCMC analysis. \textbf{b}, The posterior distribution function (PDF) of n vs coupling, shown for each NV. \textbf{c}, The exponent stretching factors are found by fitting the PDF of n to two Gaussians.}
    \label{fig:twoExpMCMC}
\end{figure}

For the shallow NV centers investigated in this work, a single exponential with $n = 2$ fixed leads to a poor fit while permitting $n$ to vary leads to better fits with $n < 2$, in general. Additionally, we claim that for configurationally averaged surface spins, the short and long time behavior of the FID should have different stretching factors. In particular, it should be possible to observe $n = 2$ for short times and $n = 2/3$ for long times in NV centers that are closer to the surface than the average surface spin separation. Here we investigate the veracity and robustness of this claim by fitting the DEER signal of four NV centers from \elanasample{}. These are the three NV centers depicted in Fig.~3 of the Main Text, plus one more that is located at a similar depth to the shallowest. We fit these data to a function of the form

\begin{equation}
\label{eq:twoexpeqn}
f(\tau) = a\exp{-(\Gamma_1 \tau)^n_1} + (1 - a)\exp{-(\Gamma_2\tau)^n_2}\,,
\end{equation}

\noindent{}where $\tau$ is the free precession time and $a, \Gamma_1, \Gamma_2, n_1, n_2$ are the fit parameters. Here we take a fully Bayesian approach and use Markov Chain Monte Carlo (MCMC) to sample the posterior distribution of these parameters, using a Gaussian likelihood with $\mu = f(\tau)$ and $\sigma(\tau)$ from the actual variance of the data. Priors for the fit parameters are uniform distributions. The resulting posterior distributions of $n_1$ and $n_2$ are shown in Fig \ref{fig:twoExpMCMC}. We find that the early time exponent, $n_2$, has a mean value close to two while the late time exponent $n_1$ is lower. For the deepest NV center, the two distributions are nearly indistinguishable, indicating that the biexponential has collapsed into a single exponential, which is expected for an NV center that is deeper than the average spin-spin separation.

\begin{figure}[H]
    \centering
    \includegraphics[width=5in]{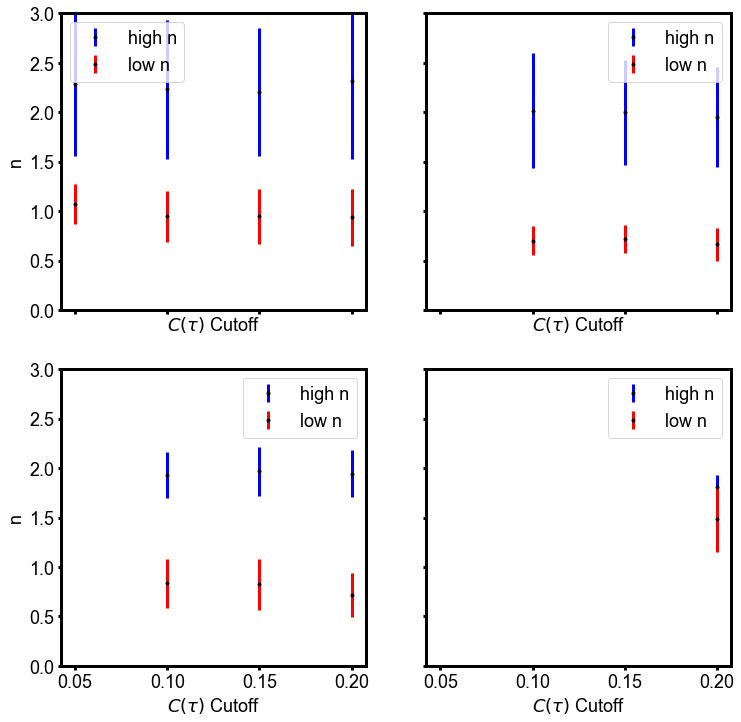}
    \caption{\textbf{Stability of stretching factors.} Mean and standard error of the two stretching factors (high $n$ and low $n$) of equation \eqref{eq:twoexpeqn} as a function of the cutoff value of the coherence for the four NV centers in Fig.~\ref{fig:twoExpMCMC}. For two NV centers, $C(\tau)$ did not dip below 0.1 during the measurement period and for one NV center, $C(\tau)$ did not dip below 0.2 during the measurement period. However, extracted exponents remained stable for all NV centers regardless of the exact $C(\tau)$ cutoff value.}
    \label{fig:varycutoff}
\end{figure}

Our data processing procedure, as described in Section \ref{sec:fittingprocedures}, necessitates choosing a cutoff value of $C(\tau)$ in time after which data is discarded. This is to prevent the signal to noise ratio of the FID data from becoming too large, which is an unavoidable consequence of dividing the DEER echo by the spin echo coherence. We show in Fig.~\ref{fig:varycutoff} that reasonable choices of the cutoff value do not affect the results of the fit. That is, for these datasets, there is enough coherence left at the cutoff point that the fit parameters are well constrained.

\subsection{Model comparison}
\label{sec:Modelcomparison}
While it is instructive to extract bi-exponential stretching factors observe short and long time behavior, Eq.~\eqref{eq:twoexpeqn} is not the form that the free induction decay is expected to take in either the static spin or configurational averaging picture. To assess the utility of the configurational averaging model, we compare the fit of the Monte Carlo simulated spin hopping model to the single exponential model fit for the four NV centers in Figure \ref{fig:twoExpMCMC} (three of which are also depicted in Fig.~3 of the Main Text)  using several standard Bayesian model comparisons \cite{vehtari2017practical, gregory2005bayesian}. For each of these NV centers, we report the Widely Applicable Information Criteria (WAIC) and associated error, $\sigma_{\rm{WAIC}}$, the Leave-One-Out (LOO) cross validation statistic and associated error, $\sigma_{\rm{LOO}}$, and the total model log-likelihood and associated error, $\sigma_{\rm{Log-likelihood}}$, for the two different models. The results are reported in Table \ref{comparisontable}, from which it is apparent that the configurational averaging model is preferred over the single exponential predicted from a static bath in all cases, and that this difference is significant for two NV centers. The reason for the low significance of NV1 is likely the lower signal to noise of this measurement, and for NV4 the two models are expected to perform similarly due to the depth of this NV center.

For the single exponential fit, the stretching factor $n$ is constrained to be $n \geq 1$ following the discussion of noise sources in Section \ref{n_derivation}. For the configurational averaging Monte Carlo model, both the depth and the surface spin density are free parameters. In order to remove any ambiguity regarding data cutoffs, the coherence and the DEER signal are fit simultaneously with separate Gaussian likelihoods. This is possible for these NV centers because the host diamond is isotopically purified, and the absence of $^{13}$C oscillations greatly simplifies the fitting. The coherence is therefore assumed to take the form of a stretched single exponential.
\begin{table}[H]
\centering

\begin{tabular}{|l|l|l|l|l|l|l|l|}
\hline
\multicolumn{1}{|c|}{\textbf{NV Center}} &
  \multicolumn{1}{c|}{\textbf{Model}} &
  \multicolumn{1}{c|}{\textbf{WAIC}} &
  \multicolumn{1}{c|}{\textbf{$\sigma_{\rm{WAIC}}$}} &
  \multicolumn{1}{c|}{\textbf{LOO}} &
  \multicolumn{1}{c|}{\textbf{$\sigma_{\rm{LOO}}$}} &
  \multicolumn{1}{c|}{\textbf{Log-likelihood}} &
  \multicolumn{1}{c|}{\textbf{$\sigma_{\rm{Log-likelihood}}$}} \\ \Xhline{4\arrayrulewidth}
\multirow{2}{*}{NV1 (3.6 nm)} & Hopping & 196 & 12 & 196 & 11 & 1243 & 13 \\ \cline{2-8}
                     & Static  & 192 & 12 & 192 & 11 & 1235 & 14 \\ \hline
\multirow{2}{*}{NV2 (5.3 nm)} & Hopping & 253 & 9  & 253 & 9  & 1429 & 15 \\ \cline{2-8}
                     & Static  & 226 & 2  & 226 & 12 & 1401 & 17 \\ \hline
\multirow{2}{*}{NV3 (6.5 nm)} & Hopping & 277 & 8  & 277 & 9  & 1543 & 15 \\ \cline{2-8}
                     & Static  & 244 & 11 & 244 & 12 & 1497 & 19 \\ \hline
\multirow{2}{*}{NV4 (9.3 nm)} & Hopping & 293 & 20 & 293 & 20 & 1664 & 18 \\ \cline{2-8}
                     & Static  & 288 & 22 & 288 & 22 & 1501 & 72 \\ \hline
\end{tabular}
\caption{Model comparison between static spins (single exponent) and configurational averaging models for four highly sampled NV center datasets.\label{comparisontable}}
\end{table}

\section{Fraction of spins that are hopping}
\label{sec:fractionhopping}

\begin{figure}[H]
\includegraphics[width=4.45in]{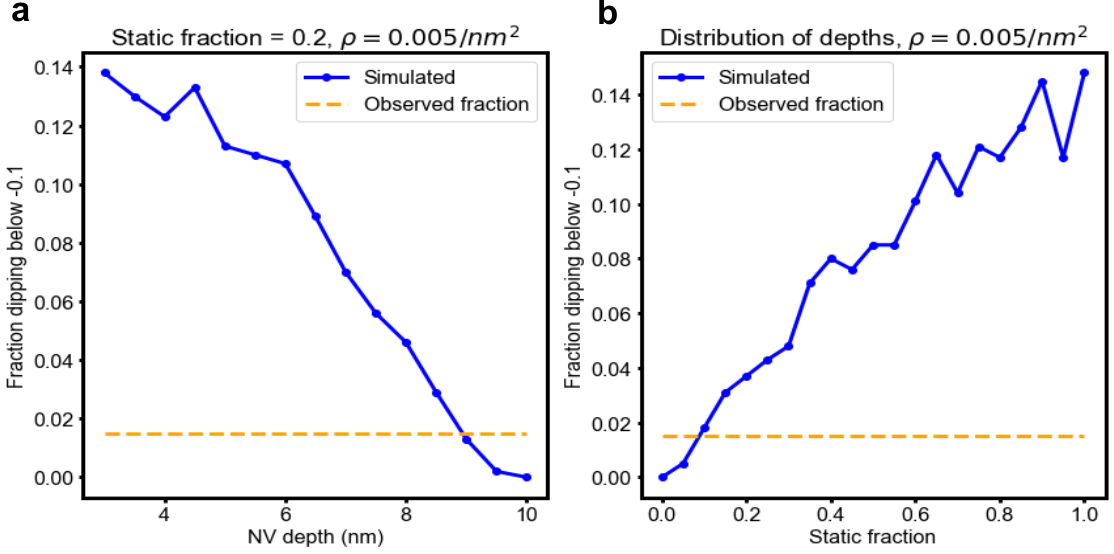}
\caption{\textbf{Stationary Surface spin fraction.} \textbf{a},  Simulation of the fraction of NV centers that exhibit strong coupling in DEER as a function of depth, if 20$\%$ of spins on the surface are placed randomly relative to the NV center and then do not hop. \textbf{b}, Simulation of the fraction of NV centers that exhibit strong coupling in their DEER signal as a function of the fraction of spins that are stationary on the surface, given a density of $\sigma = 0.005/$nm$^2$, and depths drawn from the distribution of depths measured across all samples. Orange dashed line corresponds to the observed fraction of 3 strongly coupled DEER signals out of 191 NV centers. }
\label{fig:sshop}
\end{figure}

Across the nine samples listed in Table \ref{tab:sampletable} and an additional five samples with similar implantation parameters and surface preparations, we measured DEER signals on 191 unique NV centers and observed only three spins that exhibited obvious coherent oscillations, as defined by a DEER contrast (DEER echo divided by coherence) that dipped below -0.1. This ratio allows us to place a bound on the number of spins that could still be stationary while the remaining spins hop. If we assume typical extracted densities of $\sigma = 0.005/$nm$^2$, then given that the typical NV center depths in our samples are on average less than 10 nm, we expect that less than 20$\%$ of surface spins are stationary (see Fig.~\ref{fig:sshop}).

\bibliographystyle{naturemag}
\bibliography{references.bib}

\begin{thebibliography}{10}
\expandafter\ifx\csname url\endcsname\relax
  \def\url#1{\texttt{#1}}\fi
\expandafter\ifx\csname urlprefix\endcsname\relax\def\urlprefix{URL }\fi
\providecommand{\bibinfo}[2]{#2}
\providecommand{\eprint}[2][]{\url{#2}}

\bibitem{maze2008nanoscale}
\bibinfo{author}{Maze, J.~R.} \emph{et~al.}
\newblock \bibinfo{title}{Nanoscale magnetic sensing with an individual
  electronic spin in diamond}.
\newblock \emph{\bibinfo{journal}{Nature}} \textbf{\bibinfo{volume}{455}},
  \bibinfo{pages}{644--647} (\bibinfo{year}{2008}).

\bibitem{cai2013large}
\bibinfo{author}{Cai, J.}, \bibinfo{author}{Retzker, A.},
  \bibinfo{author}{Jelezko, F.} \& \bibinfo{author}{Plenio, M.~B.}
\newblock \bibinfo{title}{A large-scale quantum simulator on a diamond surface
  at room temperature}.
\newblock \emph{\bibinfo{journal}{Nature Physics}}
  \textbf{\bibinfo{volume}{9}}, \bibinfo{pages}{168--173}
  (\bibinfo{year}{2013}).

\bibitem{schlipf2017molecular}
\bibinfo{author}{Schlipf, L.} \emph{et~al.}
\newblock \bibinfo{title}{A molecular quantum spin network controlled by a
  single qubit}.
\newblock \emph{\bibinfo{journal}{Science advances}}
  \textbf{\bibinfo{volume}{3}}, \bibinfo{pages}{e1701116}
  (\bibinfo{year}{2017}).

\bibitem{Grotz2011}
\bibinfo{author}{Grotz, B.} \emph{et~al.}
\newblock \bibinfo{title}{{Sensing external spins with nitrogen-vacancy
  diamond}}.
\newblock \emph{\bibinfo{journal}{New Journal of Physics}}
  \textbf{\bibinfo{volume}{13}}, \bibinfo{pages}{055004}
  (\bibinfo{year}{2011}).
\newblock
  \urlprefix\url{http://stacks.iop.org/1367-2630/13/i=5/a=055004?key=crossref.d3a683be647b9a50f62b6eaf6f8a7495}.

\bibitem{Grinolds2014}
\bibinfo{author}{Grinolds, M.~S.} \emph{et~al.}
\newblock \bibinfo{title}{{Subnanometre resolution in three-dimensional
  magnetic resonance imaging of individual dark spins}}.
\newblock \emph{\bibinfo{journal}{Nature Nanotechnology}}
  \textbf{\bibinfo{volume}{9}}, \bibinfo{pages}{279--284}
  (\bibinfo{year}{2014}).
\newblock \urlprefix\url{http://www.nature.com/articles/nnano.2014.30}.

\bibitem{Lovchinsky2016a}
\bibinfo{author}{Lovchinsky, I.} \emph{et~al.}
\newblock \bibinfo{title}{{Nuclear magnetic resonance detection and
  spectroscopy of single proteins using quantum logic.}}
\newblock \emph{\bibinfo{journal}{Science (New York, N.Y.)}}
  \textbf{\bibinfo{volume}{351}}, \bibinfo{pages}{836--41}
  (\bibinfo{year}{2016}).
\newblock \urlprefix\url{http://www.ncbi.nlm.nih.gov/pubmed/26847544}.

\bibitem{Sushkov2014}
\bibinfo{author}{Sushkov, A.} \emph{et~al.}
\newblock \bibinfo{title}{{Magnetic Resonance Detection of Individual Proton
  Spins Using Quantum Reporters}}.
\newblock \emph{\bibinfo{journal}{Physical Review Letters}}
  \textbf{\bibinfo{volume}{113}}, \bibinfo{pages}{197601}
  (\bibinfo{year}{2014}).
\newblock
  \urlprefix\url{https://link.aps.org/doi/10.1103/PhysRevLett.113.197601}.

\bibitem{Myers2014}
\bibinfo{author}{Myers, B.} \emph{et~al.}
\newblock \bibinfo{title}{{Probing Surface Noise with Depth-Calibrated Spins in
  Diamond}}.
\newblock \emph{\bibinfo{journal}{Physical Review Letters}}
  \textbf{\bibinfo{volume}{113}}, \bibinfo{pages}{027602}
  (\bibinfo{year}{2014}).
\newblock
  \urlprefix\url{https://link.aps.org/doi/10.1103/PhysRevLett.113.027602}.

\bibitem{Romach2015}
\bibinfo{author}{Romach, Y.} \emph{et~al.}
\newblock \bibinfo{title}{{Spectroscopy of Surface-Induced Noise Using Shallow
  Spins in Diamond}}.
\newblock \emph{\bibinfo{journal}{Physical Review Letters}}
  \textbf{\bibinfo{volume}{114}}, \bibinfo{pages}{017601}
  (\bibinfo{year}{2015}).
\newblock
  \urlprefix\url{https://link.aps.org/doi/10.1103/PhysRevLett.114.017601}.

\bibitem{sangtawesin2019origins}
\bibinfo{author}{Sangtawesin, S.} \emph{et~al.}
\newblock \bibinfo{title}{Origins of diamond surface noise probed by
  correlating single-spin measurements with surface spectroscopy}.
\newblock \emph{\bibinfo{journal}{Physical Review X}}
  \textbf{\bibinfo{volume}{9}}, \bibinfo{pages}{031052} (\bibinfo{year}{2019}).

\bibitem{Mamin2012}
\bibinfo{author}{Mamin, H.~J.}, \bibinfo{author}{Sherwood, M.~H.} \&
  \bibinfo{author}{Rugar, D.}
\newblock \bibinfo{title}{{Detecting external electron spins using
  nitrogen-vacancy centers}}.
\newblock \emph{\bibinfo{journal}{Physical Review B}}
  \textbf{\bibinfo{volume}{86}}, \bibinfo{pages}{195422}
  (\bibinfo{year}{2012}).
\newblock \urlprefix\url{https://link.aps.org/doi/10.1103/PhysRevB.86.195422}.

\bibitem{bluvstein2019extending}
\bibinfo{author}{Bluvstein, D.}, \bibinfo{author}{Zhang, Z.},
  \bibinfo{author}{McLellan, C.~A.}, \bibinfo{author}{Williams, N.~R.} \&
  \bibinfo{author}{Jayich, A. C.~B.}
\newblock \bibinfo{title}{Extending the quantum coherence of a near-surface
  qubit by coherently driving the paramagnetic surface environment}.
\newblock \emph{\bibinfo{journal}{Physical review letters}}
  \textbf{\bibinfo{volume}{123}}, \bibinfo{pages}{146804}
  (\bibinfo{year}{2019}).

\bibitem{Pham2016}
\bibinfo{author}{Pham, L.~M.} \emph{et~al.}
\newblock \bibinfo{title}{{NMR technique for determining the depth of shallow
  nitrogen-vacancy centers in diamond}}.
\newblock \emph{\bibinfo{journal}{Physical Review B}}
  \textbf{\bibinfo{volume}{93}}, \bibinfo{pages}{045425}
  (\bibinfo{year}{2016}).
\newblock \urlprefix\url{https://link.aps.org/doi/10.1103/PhysRevB.93.045425}.

\bibitem{milov1981application}
\bibinfo{author}{Milov, A.}, \bibinfo{author}{Salikhov, K.} \&
  \bibinfo{author}{Shirov, M.}
\newblock \bibinfo{title}{Application of the double resonance method to
  electron spin echo in a study of the spatial distribution of paramagnetic
  centers in solids}.
\newblock \emph{\bibinfo{journal}{Sov. Phys. Solid State}}
  \textbf{\bibinfo{volume}{23}}, \bibinfo{pages}{565--569}
  (\bibinfo{year}{1981}).

\bibitem{childress2006coherent}
\bibinfo{author}{Childress, L.} \emph{et~al.}
\newblock \bibinfo{title}{Coherent dynamics of coupled electron and nuclear
  spin qubits in diamond}.
\newblock \emph{\bibinfo{journal}{Science}} \textbf{\bibinfo{volume}{314}},
  \bibinfo{pages}{281--285} (\bibinfo{year}{2006}).

\bibitem{rowan1965electron}
\bibinfo{author}{Rowan, L.}, \bibinfo{author}{Hahn, E.} \&
  \bibinfo{author}{Mims, W.}
\newblock \bibinfo{title}{Electron-spin-echo envelope modulation}.
\newblock \emph{\bibinfo{journal}{Physical Review}}
  \textbf{\bibinfo{volume}{137}}, \bibinfo{pages}{A61} (\bibinfo{year}{1965}).

\bibitem{Kim2015}
\bibinfo{author}{Kim, M.} \emph{et~al.}
\newblock \bibinfo{title}{{Decoherence of Near-Surface Nitrogen-Vacancy Centers
  Due to Electric Field Noise}}.
\newblock \emph{\bibinfo{journal}{Physical Review Letters}}
  \textbf{\bibinfo{volume}{115}}, \bibinfo{pages}{087602}
  (\bibinfo{year}{2015}).
\newblock
  \urlprefix\url{https://link.aps.org/doi/10.1103/PhysRevLett.115.087602}.

\bibitem{deSousa2009}
\bibinfo{author}{de~Sousa, R.}
\newblock \emph{\bibinfo{title}{Electron Spin as a Spectrometer
  of Nuclear-Spin Noise and Other Fluctuations}}, \bibinfo{pages}{183--220}
  (\bibinfo{publisher}{Springer Berlin Heidelberg}, \bibinfo{address}{Berlin,
  Heidelberg}, \bibinfo{year}{2009}).
\newblock \urlprefix\url{https://doi.org/10.1007/978-3-540-79365-6_10}.

\bibitem{klauder1962spectral}
\bibinfo{author}{Klauder, J.} \& \bibinfo{author}{Anderson, P.}
\newblock \bibinfo{title}{Spectral diffusion decay in spin resonance
  experiments}.
\newblock \emph{\bibinfo{journal}{Physical Review}}
  \textbf{\bibinfo{volume}{125}}, \bibinfo{pages}{912} (\bibinfo{year}{1962}).

\bibitem{dobrovitski2008decoherence}
\bibinfo{author}{Dobrovitski, V.}, \bibinfo{author}{Feiguin, A.},
  \bibinfo{author}{Awschalom, D.} \& \bibinfo{author}{Hanson, R.}
\newblock \bibinfo{title}{Decoherence dynamics of a single spin versus spin
  ensemble}.
\newblock \emph{\bibinfo{journal}{Physical Review B}}
  \textbf{\bibinfo{volume}{77}}, \bibinfo{pages}{245212}
  (\bibinfo{year}{2008}).

\bibitem{Stanwix2010Coherence}
\bibinfo{author}{Stanwix, P.~L.} \emph{et~al.}
\newblock \bibinfo{title}{Coherence of nitrogen-vacancy electronic spin
  ensembles in diamond}.
\newblock \emph{\bibinfo{journal}{Phys. Rev. B}} \textbf{\bibinfo{volume}{82}},
  \bibinfo{pages}{201201} (\bibinfo{year}{2010}).
\newblock \urlprefix\url{https://link.aps.org/doi/10.1103/PhysRevB.82.201201}.

\bibitem{Siyushev2013Optically}
\bibinfo{author}{Siyushev, P.} \emph{et~al.}
\newblock \bibinfo{title}{Optically controlled switching of the charge state of
  a single nitrogen-vacancy center in diamond at cryogenic temperatures}.
\newblock \emph{\bibinfo{journal}{Phys. Rev. Lett.}}
  \textbf{\bibinfo{volume}{110}}, \bibinfo{pages}{167402}
  (\bibinfo{year}{2013}).
\newblock
  \urlprefix\url{https://link.aps.org/doi/10.1103/PhysRevLett.110.167402}.

\bibitem{wolters2013measurement}
\bibinfo{author}{Wolters, J.}, \bibinfo{author}{Sadzak, N.},
  \bibinfo{author}{Schell, A.~W.}, \bibinfo{author}{Schr{\"o}der, T.} \&
  \bibinfo{author}{Benson, O.}
\newblock \bibinfo{title}{Measurement of the ultrafast spectral diffusion of
  the optical transition of nitrogen vacancy centers in nano-size diamond using
  correlation interferometry}.
\newblock \emph{\bibinfo{journal}{Physical review letters}}
  \textbf{\bibinfo{volume}{110}}, \bibinfo{pages}{027401}
  (\bibinfo{year}{2013}).

\bibitem{lacelle1993typical}
\bibinfo{author}{Lacelle, S.} \& \bibinfo{author}{Tremblay, L.}
\newblock \bibinfo{title}{What is a typical dipolar coupling constant in a
  solid?}
\newblock \emph{\bibinfo{journal}{The Journal of chemical physics}}
  \textbf{\bibinfo{volume}{98}}, \bibinfo{pages}{3642--3649}
  (\bibinfo{year}{1993}).

\bibitem{fel1996configurational}
\bibinfo{author}{Fel’dman, E.~B.} \& \bibinfo{author}{Lacelle, S.}
\newblock \bibinfo{title}{Configurational averaging of dipolar interactions in
  magnetically diluted spin networks}.
\newblock \emph{\bibinfo{journal}{The Journal of chemical physics}}
  \textbf{\bibinfo{volume}{104}}, \bibinfo{pages}{2000--2009}
  (\bibinfo{year}{1996}).

\bibitem{hahn2019long}
\bibinfo{author}{Hahn, W.} \& \bibinfo{author}{Dobrovitski, V.}
\newblock \bibinfo{title}{Long-lived coherence in driven spin systems: from
  two-to infinite spatial dimensions}.
\newblock \emph{\bibinfo{journal}{arXiv preprint arXiv:1911.06272}}
  (\bibinfo{year}{2019}).

\bibitem{liu2009photochemical}
\bibinfo{author}{Liu, H.} \emph{et~al.}
\newblock \bibinfo{title}{Photochemical reactivity of graphene}.
\newblock \emph{\bibinfo{journal}{Journal of the American Chemical Society}}
  \textbf{\bibinfo{volume}{131}}, \bibinfo{pages}{17099--17101}
  (\bibinfo{year}{2009}).

\bibitem{yazyev2007defect}
\bibinfo{author}{Yazyev, O.~V.} \& \bibinfo{author}{Helm, L.}
\newblock \bibinfo{title}{Defect-induced magnetism in graphene}.
\newblock \emph{\bibinfo{journal}{Physical Review B}}
  \textbf{\bibinfo{volume}{75}}, \bibinfo{pages}{125408}
  (\bibinfo{year}{2007}).

\bibitem{robledo2011high}
\bibinfo{author}{Robledo, L.} \emph{et~al.}
\newblock \bibinfo{title}{High-fidelity projective read-out of a solid-state
  spin quantum register}.
\newblock \emph{\bibinfo{journal}{Nature}} \textbf{\bibinfo{volume}{477}},
  \bibinfo{pages}{574--578} (\bibinfo{year}{2011}).

\bibitem{irber2020robust}
\bibinfo{author}{Irber, D.~M.} \emph{et~al.}
\newblock \bibinfo{title}{Robust all-optical single-shot readout of nv centers
  in diamond} (\bibinfo{year}{2020}).
\newblock \eprint{2006.02938}.

\bibitem{lovchinsky2016nuclear}
\bibinfo{author}{Lovchinsky, I.} \emph{et~al.}
\newblock \bibinfo{title}{Nuclear magnetic resonance detection and spectroscopy
  of single proteins using quantum logic}.
\newblock \emph{\bibinfo{journal}{Science}} \textbf{\bibinfo{volume}{351}},
  \bibinfo{pages}{836--841} (\bibinfo{year}{2016}).

\bibitem{grinolds2011quantum}
\bibinfo{author}{Grinolds, M.} \emph{et~al.}
\newblock \bibinfo{title}{Quantum control of proximal spins using nanoscale
  magnetic resonance imaging}.
\newblock \emph{\bibinfo{journal}{Nature Physics}}
  \textbf{\bibinfo{volume}{7}}, \bibinfo{pages}{687--692}
  (\bibinfo{year}{2011}).

\bibitem{Rugar2004}
\bibinfo{author}{Rugar, D.}, \bibinfo{author}{Budakian, R.},
  \bibinfo{author}{Mamin, H.~J.} \& \bibinfo{author}{Chui, B.~W.}
\newblock \bibinfo{title}{{Single spin detection by magnetic resonance force
  microscopy}}.
\newblock \emph{\bibinfo{journal}{Nature}} \textbf{\bibinfo{volume}{430}},
  \bibinfo{pages}{329--332} (\bibinfo{year}{2004}).
\newblock \urlprefix\url{http://www.nature.com/articles/nature02658}.

\bibitem{maletinsky2012robust}
\bibinfo{author}{Maletinsky, P.} \emph{et~al.}
\newblock \bibinfo{title}{A robust scanning diamond sensor for nanoscale
  imaging with single nitrogen-vacancy centres}.
\newblock \emph{\bibinfo{journal}{Nature nanotechnology}}
  \textbf{\bibinfo{volume}{7}}, \bibinfo{pages}{320--324}
  (\bibinfo{year}{2012}).

\bibitem{Pelliccione2016b}
\bibinfo{author}{Pelliccione, M.} \emph{et~al.}
\newblock \bibinfo{title}{{Scanned probe imaging of nanoscale magnetism at
  cryogenic temperatures with a single-spin quantum sensor}}.
\newblock \emph{\bibinfo{journal}{Nature Nanotechnology}}
  \textbf{\bibinfo{volume}{11}}, \bibinfo{pages}{700--705}
  (\bibinfo{year}{2016}).
\newblock \urlprefix\url{http://www.nature.com/articles/nnano.2016.68}.

\bibitem{davis}
\bibinfo{author}{Davis, E.~J.} \emph{et~al.}
\newblock \bibinfo{title}{Probing many-body noise in a strongly-interacting
  two-dimensional dipolar spin system}.
\newblock \emph{\bibinfo{journal}{Manuscript in preparation}}
  (\bibinfo{year}{2021}).

\bibitem{stohr1992springer}
\bibinfo{author}{St{\"o}hr, J.}
\newblock \bibinfo{title}{Springer series in surface science}.
\newblock \emph{\bibinfo{journal}{NEXAFS spectroscopy}}
  \textbf{\bibinfo{volume}{25}} (\bibinfo{year}{1992}).

\bibitem{bobrov2001electronic}
\bibinfo{author}{Bobrov, K.}, \bibinfo{author}{Comtet, G.},
  \bibinfo{author}{Dujardin, G.} \& \bibinfo{author}{Hellner, L.}
\newblock \bibinfo{title}{Electronic structure of partially hydrogenated si
  (100)-(2$\times$ 1) surfaces prepared by thermal and nonthermal desorption}.
\newblock \emph{\bibinfo{journal}{Physical review letters}}
  \textbf{\bibinfo{volume}{86}}, \bibinfo{pages}{2633} (\bibinfo{year}{2001}).

\bibitem{wang2012comparison}
\bibinfo{author}{Wang, Z.-H.}, \bibinfo{author}{De~Lange, G.},
  \bibinfo{author}{Rist{\`e}, D.}, \bibinfo{author}{Hanson, R.} \&
  \bibinfo{author}{Dobrovitski, V.}
\newblock \bibinfo{title}{Comparison of dynamical decoupling protocols for a
  nitrogen-vacancy center in diamond}.
\newblock \emph{\bibinfo{journal}{Physical Review B}}
  \textbf{\bibinfo{volume}{85}}, \bibinfo{pages}{155204}
  (\bibinfo{year}{2012}).

\bibitem{kaviani2014proper}
\bibinfo{author}{Kaviani, M.} \emph{et~al.}
\newblock \bibinfo{title}{Proper surface termination for luminescent
  near-surface nv centers in diamond}.
\newblock \emph{\bibinfo{journal}{Nano letters}} \textbf{\bibinfo{volume}{14}},
  \bibinfo{pages}{4772--4777} (\bibinfo{year}{2014}).

\bibitem{chadi1987stabilities}
\bibinfo{author}{Chadi, D.}
\newblock \bibinfo{title}{Stabilities of single-layer and bilayer steps on si
  (001) surfaces}.
\newblock \emph{\bibinfo{journal}{Physical Review Letters}}
  \textbf{\bibinfo{volume}{59}}, \bibinfo{pages}{1691} (\bibinfo{year}{1987}).

\bibitem{yang2008bond}
\bibinfo{author}{Yang, H.}, \bibinfo{author}{Xu, L.}, \bibinfo{author}{Fang,
  Z.}, \bibinfo{author}{Gu, C.} \& \bibinfo{author}{Zhang, S.}
\newblock \bibinfo{title}{Bond-counting rule for carbon and its application to
  the roughness of diamond (001)}.
\newblock \emph{\bibinfo{journal}{Physical review letters}}
  \textbf{\bibinfo{volume}{100}}, \bibinfo{pages}{026101}
  (\bibinfo{year}{2008}).

\bibitem{kresse1996efficient}
\bibinfo{author}{Kresse, G.} \& \bibinfo{author}{Furthm{\"u}ller, J.}
\newblock \bibinfo{title}{Efficient iterative schemes for ab initio
  total-energy calculations using a plane-wave basis set}.
\newblock \emph{\bibinfo{journal}{Physical review B}}
  \textbf{\bibinfo{volume}{54}}, \bibinfo{pages}{11169} (\bibinfo{year}{1996}).

\bibitem{blochl1994projector}
\bibinfo{author}{Bl{\"o}chl, P.~E.}
\newblock \bibinfo{title}{Projector augmented-wave method}.
\newblock \emph{\bibinfo{journal}{Physical review B}}
  \textbf{\bibinfo{volume}{50}}, \bibinfo{pages}{17953} (\bibinfo{year}{1994}).

\bibitem{perdew1996generalized}
\bibinfo{author}{Perdew, J.~P.}, \bibinfo{author}{Burke, K.} \&
  \bibinfo{author}{Wang, Y.}
\newblock \bibinfo{title}{Generalized gradient approximation for the
  exchange-correlation hole of a many-electron system}.
\newblock \emph{\bibinfo{journal}{Physical Review B}}
  \textbf{\bibinfo{volume}{54}}, \bibinfo{pages}{16533} (\bibinfo{year}{1996}).

\bibitem{heyd2003hybrid}
\bibinfo{author}{Heyd, J.}, \bibinfo{author}{Scuseria, G.~E.} \&
  \bibinfo{author}{Ernzerhof, M.}
\newblock \bibinfo{title}{Hybrid functionals based on a screened coulomb
  potential}.
\newblock \emph{\bibinfo{journal}{The Journal of chemical physics}}
  \textbf{\bibinfo{volume}{124}} (\bibinfo{year}{2006}).

\bibitem{gali2009theory}
\bibinfo{author}{Gali, A.}, \bibinfo{author}{Janz{\'e}n, E.},
  \bibinfo{author}{De{\'a}k, P.}, \bibinfo{author}{Kresse, G.} \&
  \bibinfo{author}{Kaxiras, E.}
\newblock \bibinfo{title}{Theory of spin-conserving excitation of the n- v-
  center in diamond}.
\newblock \emph{\bibinfo{journal}{Physical review letters}}
  \textbf{\bibinfo{volume}{103}}, \bibinfo{pages}{186404}
  (\bibinfo{year}{2009}).

\bibitem{laraoui2013high}
\bibinfo{author}{Laraoui, A.} \emph{et~al.}
\newblock \bibinfo{title}{High-resolution correlation spectroscopy of 13 c
  spins near a nitrogen-vacancy centre in diamond}.
\newblock \emph{\bibinfo{journal}{Nature communications}}
  \textbf{\bibinfo{volume}{4}}, \bibinfo{pages}{1--7} (\bibinfo{year}{2013}).

\bibitem{Slichter1990}
\bibinfo{author}{Slichter, C.~P.}
\newblock \emph{\bibinfo{title}{{Principles of magnetic resonance}}}
  (\bibinfo{publisher}{Springer-Verlag}, \bibinfo{year}{1990}).

\bibitem{Abragam1961}
\bibinfo{author}{Abragam, A.}
\newblock \emph{\bibinfo{title}{{The principles of nuclear magnetism}}}
  (\bibinfo{publisher}{Oxford university press}, \bibinfo{year}{1961}).

\bibitem{Hall2016}
\bibinfo{author}{Hall, L.~T.} \emph{et~al.}
\newblock \bibinfo{title}{{Detection of nanoscale electron spin resonance
  spectra demonstrated using nitrogen-vacancy centre probes in diamond}}.
\newblock \emph{\bibinfo{journal}{Nature Communications}}
  \textbf{\bibinfo{volume}{7}}, \bibinfo{pages}{10211} (\bibinfo{year}{2016}).
\newblock \urlprefix\url{http://www.nature.com/articles/ncomms10211}.

\bibitem{vehtari2017practical}
\bibinfo{author}{Vehtari, A.}, \bibinfo{author}{Gelman, A.} \&
  \bibinfo{author}{Gabry, J.}
\newblock \bibinfo{title}{Practical bayesian model evaluation using
  leave-one-out cross-validation and waic}.
\newblock \emph{\bibinfo{journal}{Statistics and computing}}
  \textbf{\bibinfo{volume}{27}}, \bibinfo{pages}{1413--1432}
  (\bibinfo{year}{2017}).

\bibitem{gregory2005bayesian}
\bibinfo{author}{Gregory, P.}
\newblock \emph{\bibinfo{title}{Bayesian logical data analysis for the physical
  sciences: a comparative approach with mathematica{\textregistered} support}}
  (\bibinfo{publisher}{Cambridge University Press}, \bibinfo{year}{2005}).

\end{thebibliography}

\end{document}